\documentclass[aps,prb,showpacs]{revtex4}

\usepackage{amsmath,amssymb,mathrsfs}
\usepackage[dvips]{graphicx}

\newcommand{\comm}[2]{\left[#1,#2\right]}

\newcommand{\ket}[1]{\left|#1\right\rangle}

\newcommand{\up}{\uparrow}
\newcommand{\dw}{\downarrow}

\def\ie{\emph{i.e.},\ }
\def\eg{\emph{e.g.}\ }
\def\ea{\emph{et al.}\ }
\def\b{{\mathrm{b}}}

\def\m{{\mathrm{m}}}
\def\c{{\mathrm{c}}}
\def\y{{\mathrm{y}}}

\begin{document}
\title{Many-spinon states and representations
  of Yangians in the\\ SU(\textit{n}) Haldane--Shastry model}
\author{Dirk Schuricht} 
\affiliation{The  Rudolf Peierls Centre for Theoretical Physics, 
  University of Oxford,\\ 1 Keble Road, Oxford OX1 3NP, United Kingdom} 
\begin{abstract}
  We study the relation of Yangians and spinons in the SU($n$)
  Haldane--Shastry model. The representation theory of the Yangian is shown
  to be intimately related to the fractional statistics of the spinons.  We
  construct the spinon Hilbert space from tensor products of the fundamental
  representations of the Yangian.
\end{abstract}
\date{November 9, 2007}
\maketitle

\section{Introduction}

Quantum groups~\cite{Drinfeld85,Jimbo85} first arose from the quantum inverse
scattering method~\cite{Faddeev82,KorepinBogoliubovIzergin93}, which had been
developed to construct and solve integrable quantum systems.  In particular,
quantum groups provide a way to construct and study the solutions, called
R-matrices, of the quantum Yang--Baxter equation.  Mathematically, quantum
groups are deformations of the universal enveloping algebra of the classical
Lie algebras. In general, they depend on a parameter $h$ and the underlying
Lie algebra is recovered in the limit $h\rightarrow 0$.  Yangians are special
quantum groups which were first introduced by Drinfel'd~\cite{Drinfeld85} in
1985.  Their representation theory is intimately
related~\cite{ChariPressley90,ChariPressley98} to the rational R-matrices.

Later it was discovered that Yangians also appear as additional symmetries of
quantum field theories~\cite{Bernard91,Haldane-92}, and furthermore, that
Yangians are part of the symmetry algebra of special integrable spin systems.
In particular, the one-dimensional nearest-neighbour Heisenberg model
possesses a Yangian symmetry in the limit of a chain of infinite
length~\cite{Bernard93}, whereas the Haldane--Shastry model possesses a
Yangian symmetry even for a chain of finite
length~\cite{Haldane-92,Hikami95npb}. In addition, a Yangian
symmetry exists for the one-dimensional Hubbard model on an infinite
chain~\cite{UglovKorepin94} as well as for a finite chain with suitable
hopping
amplitudes~\cite{GoehmannInozemtsev96,EsslerFrahmGoehmannKluemperKorepin05}.

From a physical point of view, the Haldane--Shastry model
(HSM)~\cite{Haldane88} owes its special importance to two reasons.  The first
and more technical one is that the model is exactly solvable even for a chain
of finite length. It is possible to derive explicit wave functions for the
ground state and the elementary spinon excitations~\cite{Haldane91prl1}.  The
second and more fundamental reason is that the HSM possesses non-interacting
or free spinon excitations~\cite{Haldane94}, a conclusion which is in
particular supported by the trivial spinon-spinon scattering matrix calculated
by Essler~\cite{Essler95} using the asymptotic Bethe Ansatz.  In 2001 this
picture was challenged by Bernevig~\emph{et al.}~\cite{Bernevig-01prb}, who
studied the explicit two-spinon wave functions and claimed to have identified
effects of an interaction between the spinons.  A critical
re-examination~\cite{GS05prb} of their conclusions, however, showed that these
alleged interaction effects are in fact due to the fractional statistics of
the spinons~\cite{Haldane91prl2}, which results in non-trivial quantisation
rules for the individual spinon momenta~\cite{Greiter}.  This debate showed
that free particles may appear interacting at first sight if an inappropriate
representation is chosen.

In this article we investigate the relation between the Yangian symmetry and
the physical properties of the spinons. We show that individual spinons in the
HSM transform under the fundamental representation of the Yangian.  We then
study the implications of the Yangian symmetry on many-spinon states.  The
main result of this analysis is the derivation of a general rule governing the
fractional statistics of the spinons.  This rule states that in the spinon
Hilbert space only the irreducible subrepresentations of the tensor products
of certain fundamental representations of the Yangian exists.  This enables us
to derive, starting from a set of individual spinon momenta, the allowed
values of the total spin of the corresponding many-spinon states, which are
subject to highly non-trivial restrictions due to the fractional statistics of
the spinons.  All results are generalised to the elementary excitations of the
SU(3) HSM.

Before we discuss the main topic of this article, we will briefly
review the HSM and its most important physical features, and give a
concise introduction to Yangians and their representation theory.

\section{Haldane--Shastry model}\label{sec:hsm}

In 1988 Haldane and Shastry discovered independently~\cite{Haldane88} that a
trial wave function proposed by Gutzwiller~\cite{Gutzwiller63} in 1963
provides the exact ground state to a Heisenberg type spin Hamiltonian whose
interaction strength falls off as the inverse square of the distance between
two spins on the chain.  Later the model was generalised to an SU($n$) spins
by Kawakami~\cite{Kawakami92prb2}.

\begin{figure}[t]
\begin{center}
\setlength{\unitlength}{10pt}
\begin{picture}(32,11)(1,0)
\put(1,10){a)}
\put(12,10){\textbf{3}}
\put(2,5){\line(1,0){10}}
\put(7,0){\line(0,1){10}}
\put(12,4.3){$J^3$}
\put(7.2,9.6){$J^8$}
\put(7,1.5){\circle*{0.5}}
\put(6,1.5){g}
\put(9.93,6.75){\circle*{0.5}}
\put(10.43,6.75){b}
\put(4.07,6.75){\circle*{0.5}}
\put(4.57,6.75){r}
\put(6.8,6.75){\rule{4.8pt}{0.3pt}}
\put(7.3,6.5){$\textstyle\frac{1}{2\sqrt{3}}$}
\put(7.3,1.25){$\textstyle\frac{-1}{\sqrt{3}}$}
\put(4.07,4.8){\rule{0.3pt}{4.8pt}}
\put(3.06,3.8){$-\textstyle\frac{1}{2}$}
\put(9.93,4.8){\rule{0.3pt}{4.8pt}}
\put(9.7,3.8){$\textstyle\frac{1}{2}$}

\put(19,10){b)}
\put(30,10){$\boldsymbol{\bar{3}}$}
\put(20,5){\line(1,0){10}}
\put(25,0){\line(0,1){10}}
\put(30,4.3){$J^3$}
\put(25.2,9.6){$J^8$}
\put(25,8.5){\circle*{0.5}}
\put(23.7,8.5){m}
\put(27.93,3.25){\circle*{0.5}}
\put(28.43,2.75){c}
\put(22.07,3.25){\circle*{0.5}}
\put(22.57,2.75){y}
\put(24.8,3.25){\rule{4.8pt}{0.3pt}}
\put(25.3,3){$\textstyle\frac{-1}{2\sqrt{3}}$}
\put(25.3,8.25){$\textstyle\frac{1}{\sqrt{3}}$}
\put(22.07,4.8){\rule{0.3pt}{4.8pt}}
\put(21.06,5.7){$-\textstyle\frac{1}{2}$}
\put(27.93,4.8){\rule{0.3pt}{4.8pt}}
\put(27.7,5.7){$\textstyle\frac{1}{2}$}
\end{picture}
\end{center}
\caption{Weight diagrams of the three-dimensional representations of
  SU(3). $J^3$ and $J^8$ span the Cartan subalgebra~\cite{Cornwell84vol2} of
  su(3).}
\label{fig:weightdiagrams}
\end{figure}
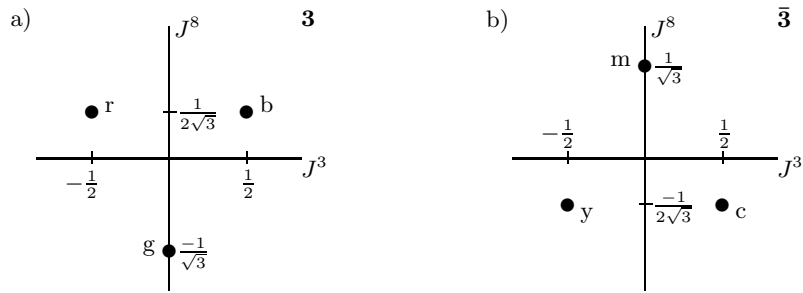

The HSM is most conveniently formulated by embedding the
one-dimensional chain with periodic boundary conditions into the
complex plane by mapping it onto the unit circle with the (SU($n$)) spins
located at complex positions
$\eta_\alpha=\exp\!\left(i\frac{2\pi}{N}\alpha\right)$, where $N$
denotes the number of sites and $\alpha=1,\ldots,N$.  The Hamiltonian
is given by~\cite{Haldane88}
\begin{equation}
  \label{eq:ham}
  H_{\mathrm{HS}}
  =\frac{2\pi^2}{N^2}
  \sum^N_{\alpha\neq\beta}
  \frac{\boldsymbol{S}_{\alpha}\!\cdot\!\boldsymbol{S}_{\beta}}
  {\vert \eta_{\alpha}-\eta_{\beta}\vert^2}.
\end{equation}
The SU(3) HSM~\cite{Kawakami92prb2} is given by replacing
$\boldsymbol{S}_\alpha$ by the eight-dimensional SU(3) spin vector
$\boldsymbol{J}_{\alpha}=\frac{1}{2}\sum_{\sigma\tau}
c_{\alpha\sigma}^{\dagger} \boldsymbol{\lambda}_{\sigma\tau}
c_{\alpha\tau}^{\phantom{\dagger}}$, where $\boldsymbol{\lambda}$ a vector
consisting of the eight Gell-Mann matrices~\cite{Cornwell84vol2}, and $\sigma$
and $\tau$ are SU(3) spin or colour indices which take the values blue (b),
red (r), or green (g) (see Fig.~\ref{fig:weightdiagrams}.a). The spins on the
lattice sites transform under the fundamental representation $\boldsymbol{n}$
of SU($n$), \eg $S=1/2$ for SU(2).

The ground state ($N=2M$, $M$ integer) of the SU(2) model is given by
\begin{equation}
  \label{eq:groundstate}
  \ket{\Psi_0}=P_{\mathrm{G}}\ket{\Psi_{\mathrm{SD}}^N},\quad
  \ket{\Psi_{\mathrm{SD}}^N}\equiv
  \prod_{q\in\mathcal{I}} c_{q\up}^\dagger c_{q\dw}^\dagger\ket{0},
\end{equation}
where the Gutzwiller projector $P_{\mathrm{G}}$ eliminates configurations with
more than one particle on any site and the interval $\mathcal{I}$ contains $M$
adjacent momenta.  For SU($n$), each momentum in $\mathcal{I}$ has to be
occupied by $n$ particles with different
spins~\cite{Kawakami92prb3}.

The model is invariant under global SU(2) or SU(3) rotations generated by
\begin{equation}
\boldsymbol{S}=\sum_{\alpha=1}^N\boldsymbol{S}_{\alpha}
\quad\mathrm{(for\ SU(2))}
\quad\mathrm{or}\quad
\boldsymbol{J}=\sum_{\alpha=1}^N\boldsymbol{J}_\alpha
\quad\mathrm{(for\ SU(3))},
\label{eq:su3-Jsymmetry}
\end{equation}
respectively.  The system possesses an additional
symmetry~\cite{Haldane-92,Hikami95npb}, which is given by
\begin{equation}
  \boldsymbol{\Lambda}=\frac{\mathrm{i}}{2}\sum_{\alpha\neq\beta}^N\,
  \frac{\eta_\alpha + \eta_\beta}{\eta_\alpha - \eta_\beta}\,
  (\boldsymbol{S}_\alpha\times\boldsymbol{S}_\beta)\quad\mathrm{or}\quad
\Lambda^a=\frac{1}{2}\sum^N_{\alpha\neq\beta}
\frac{\eta_\alpha +\eta_\beta}{\eta_\alpha -\eta_\beta}\;
f^{abc}J^b_\alpha J^c_\beta,
\label{eq:su3-Upsilon}
\end{equation}
(we use the Einstein summation convention) where $a,b,c=1,\ldots,8$ and
$f^{abc}$ denote the structure constants of SU(3).  The total spin
(\ref{eq:su3-Jsymmetry}) and rapidity operator (\ref{eq:su3-Upsilon}) generate
the Yangian, which we will discuss in detail below.

The elementary excitations of the SU($n$) HSM are constructed by annihilation
of a particle from the Slater determinant state before Gutzwiller
projection~\cite{Haldane91prl1,SG05epl},
\begin{equation}
  \ket{\Psi_{p\bar\sigma}}=P_{\mathrm{G}}\;\!
  c_{-p\sigma}^{\phantom{\dagger}}\!\ket{\Psi_{\mathrm{SD}}^{N+1}},
  \quad N=nM-1.
  \label{eq:loccoloron}
\end{equation}
Here $p$ denotes the momentum, $\sigma$ either the spin (for $n=2$) or one of
the colons blue, red or green (for $n=3$). In order to ensure that every site
is occupied by a spin after the projection, we annihilate a particle from the
Slater determinant state with $N+1$ particles.  Note that for SU(2) the
annihilation of an up-spin electron creates a down-spin spinon and vice versa.
The spinons possess spin $1/2$ like the electrons on the lattice sites.  For
SU(3) the situation is, however, fundamentally different.  Here, the
annihilation of a, say, blue particle creates an anti-blue SU(3) spinon or
coloron.  (We will use the terms SU(3) spinon and coloron simultaneously.)
This means that colorons transform under the conjugate representation
$\boldsymbol{\bar{3}}$ (see Fig.~\ref{fig:weightdiagrams}.b), if the particles
on the sites transform under
$\boldsymbol{3}$~\cite{BouwknegtSchoutens96,SG05epl}.

A non-orthogonal but complete basis for spin-polarised two-spinon eigenstates
with total momentum $p=-k_1-k_2$ is given by (we assume $k_1>k_2$)
\begin{equation}
  \ket{\Psi_{p_1\bar{\sigma},p_2\bar{\sigma}}}=
  P_{\mathrm{G}}\;\!c_{k_1\sigma}c_{k_2\sigma}
  \!\ket{\Psi_{\mathrm{SD}}^{N+2}},
  \quad N=nM-2.
  \label{eq:twospinons}
\end{equation}
These states are not energy eigenstates, but as $H_{\mathrm{HS}}$ scatters
$k_1$ and $k_2$ in only one direction (increasing $k_1-k_2$), there is a
one-to-one correspondence between these basis states and the exact eigenstates
of the HSM.  The total energy of the eigenstates takes the form of free
particles if and only if the single-spinon momenta are shifted with respect to
$k_{1,2}$~\cite{GS05prb,SG05epl}: 
\begin{equation}
  \label{eq:twospinonmomenta}
  p_{1,2}=-k_{1,2}\pm\frac{1}{2n}\frac{2\pi}{N},\quad p_1<p_2.
\end{equation}
The shift is due to the fractional statistics of the
spinons~\cite{Haldane91prl2,Greiter}.  In general, all two-spinon states with
the same single-spinon momenta are obtained by acting with the total spin and
rapidity operators on the polarised states (\ref{eq:twospinons}). In
particular, for SU(2) the action of $\Lambda^zS^-$ yields the two-spinon
singlet states.  However, this two-spinon singlet state
$\Lambda^zS^-\ket{\Psi_{p_1\up,p_2\up}}$ exists only for
$p_2-p_1>\frac{1}{2}\frac{2\pi}{N}$, as (\ref{eq:twospinons}) is
annihilated~\cite{Haldane-92} by $\Lambda^zS^-$ for
$p_2-p_1=\frac{1}{2}\frac{2\pi}{N}$.  For general many-spinon states or
spinons in the SU($n$) chain these restrictions on the possible values of the
total spin become highly non-trivial.

\section{Tableau formalism}\label{sec:yt}

Recently, Greiter and I introduced a formalism to obtain all existing
many-spinon states starting from a given set of single-spinon
momenta~\cite{GS07}. The formalism works as follows. To begin with, the
Hilbert space of a system of $N$ identical SU($n$) spins can be decomposed
into representations of the total spin (\ref{eq:su3-Jsymmetry}), which
commutes with (\ref{eq:ham}) and hence can be used to classify the energy
eigenstates.  These representations are compatible with the representations of
the symmetric group S$_N$ of $N$ elements, which may be expressed in terms of
Young tableaux~\cite{InuiTanabeOnodera96}.  In order to obtain these Young
tableaux, we draw for each of the $N$ spins a box numbered consecutively from
left to right.  The representations of SU($n$) are constructed by putting the
boxes together such that the numbers assigned to them increase in each row
from left to right and in each column from top to bottom.  Each tableau
obtained in this way represents an irreducible representation of SU($n$), it
further indicates symmetrisation over all boxes in the same row, and
antisymmetrisation over all boxes in the same column.  This implies that we
cannot have more than $n$ boxes on top of each other for SU($n$) spins.

\begin{figure}[tb]
\setlength{\unitlength}{10pt}
\begin{center}
\begin{picture}(31,5)(1,0)
\linethickness{0.3pt}
\put(0.2,3){\makebox(2,1){rep}}
\put(7,3){\makebox(3,1){$S_\mathrm{tot}$}}
\put(17.8,3){\makebox(7,1){$L$}}
\put(25.2,3){\makebox(5,1){$a_1,\ldots,a_L$}}

\multiput(0,0)(0,1){3}{\line(1,0){3}}
\multiput(0,0)(1,0){4}{\line(0,1){2}}
\put(0,1){\makebox(1,1){1}}
\put(0,0){\makebox(1,1){2}}
\put(1,1){\makebox(1,1){3}}
\put(1,0){\makebox(1,1){4}}
\put(2,1){\makebox(1,1){5}}
\put(2,0){\makebox(1,1){6}}
\put(8,1){\makebox(1,1){0}}
\put(10.5,0.9){\makebox(1,1){$\rightarrow$}}
\multiput(13,0)(0,1){3}{\line(1,0){3}}
\multiput(13,0)(1,0){4}{\line(0,1){2}}
\put(13,1){\makebox(1,1){1}}
\put(13,0){\makebox(1,1){2}}
\put(14,1){\makebox(1,1){3}}
\put(14,0){\makebox(1,1){4}}
\put(15,1){\makebox(1,1){5}}
\put(15,0){\makebox(1,1){6}}
\put(21,1){\makebox(1,1){0}}
\put(24,1.5){\line(1,0){7}}
\multiput(25,1.35)(1,0){6}{\line(0,1){0.3}}
\end{picture}

\begin{picture}(31,3)(1,0)
\linethickness{0.3pt}
\multiput(0,0)(0,1){3}{\line(1,0){3}}
\multiput(0,0)(1,0){4}{\line(0,1){2}}
\put(0,1){\makebox(1,1){1}}
\put(0,0){\makebox(1,1){2}}
\put(1,1){\makebox(1,1){3}}
\put(2,1){\makebox(1,1){4}}
\put(1,0){\makebox(1,1){5}}
\put(2,0){\makebox(1,1){6}}
\put(8,1){\makebox(1,1){0}}
\put(10.5,0.9){\makebox(1,1){$\rightarrow$}}
\put(13,2){\line(1,0){3}}
\put(13,1){\line(1,0){4}}
\put(13,0){\line(1,0){1}}
\put(15,0){\line(1,0){2}}
\multiput(13,1)(1,0){4}{\line(0,1){1}}
\multiput(13,0)(1,0){5}{\line(0,1){1}}
\put(13,1){\makebox(1,1){1}}
\put(13,0){\makebox(1,1){2}}
\put(14,1){\makebox(1,1){3}}
\put(15,1){\makebox(1,1){4}}
\put(15,0){\makebox(1,1){5}}
\put(16,0){\makebox(1,1){6}}
\multiput(14.5,0.5)(2,1){2}{\circle*{0.4}}
\put(21,1){\makebox(1,1){2}}
\put(24,1.5){\line(1,0){7}}
\multiput(25,1.35)(1,0){6}{\line(0,1){0.3}}
\multiput(27,1.5)(3,0){2}{\circle*{0.4}}
\put(26.5,0){\makebox(1,1){3}}
\put(29.5,0){\makebox(1,1){6}}
\end{picture}

\begin{picture}(31,3)(1,0)
\linethickness{0.3pt}
\multiput(0,1)(0,1){2}{\line(1,0){4}}
\put(0,0){\line(1,0){2}}
\multiput(0,1)(1,0){5}{\line(0,1){1}}
\multiput(0,0)(1,0){3}{\line(0,1){1}}
\put(0,1){\makebox(1,1){1}}
\put(0,0){\makebox(1,1){2}}
\put(1,1){\makebox(1,1){3}}
\put(2,1){\makebox(1,1){4}}
\put(1,0){\makebox(1,1){5}}
\put(3,1){\makebox(1,1){6}}
\put(8,1){\makebox(1,1){1}}
\put(10.5,0.9){\makebox(1,1){$\rightarrow$}}
\multiput(13,1)(0,1){2}{\line(1,0){4}}
\multiput(13,0)(2,0){2}{\line(1,0){1}}
\multiput(13,1)(1,0){5}{\line(0,1){1}}
\multiput(13,0)(1,0){4}{\line(0,1){1}}
\put(13,1){\makebox(1,1){1}}
\put(13,0){\makebox(1,1){2}}
\put(14,1){\makebox(1,1){3}}
\put(15,1){\makebox(1,1){4}}
\put(15,0){\makebox(1,1){5}}
\put(16,1){\makebox(1,1){6}}
\multiput(14.5,0.5)(2,0){2}{\circle*{0.4}}
\put(21,1){\makebox(1,1){2}}
\put(24,1.5){\line(1,0){7}}
\multiput(25,1.35)(1,0){6}{\line(0,1){0.3}}
\multiput(27,1.5)(3,0){2}{\circle*{0.4}}
\put(26.5,0){\makebox(1,1){3}}
\put(29.5,0){\makebox(1,1){6}}
\end{picture}

\begin{picture}(31,3)(1,0)
\linethickness{0.3pt}
\multiput(0,1)(0,1){2}{\line(1,0){4}}
\put(0,0){\line(1,0){2}}
\multiput(0,1)(1,0){5}{\line(0,1){1}}
\multiput(0,0)(1,0){3}{\line(0,1){1}}
\put(0,1){\makebox(1,1){1}}
\put(1,1){\makebox(1,1){2}}
\put(2,1){\makebox(1,1){3}}
\put(0,0){\makebox(1,1){4}}
\put(1,0){\makebox(1,1){5}}
\put(3,1){\makebox(1,1){6}}
\put(8,1){\makebox(1,1){1}}
\put(10.5,0.9){\makebox(1,1){$\rightarrow$}}
\put(13,2){\line(1,0){3}}
\put(17,2){\line(1,0){1}}
\put(13,1){\line(1,0){5}}
\put(15,0){\line(1,0){2}}
\multiput(13,1)(1,0){6}{\line(0,1){1}}
\multiput(15,0)(1,0){3}{\line(0,1){1}}
\put(13,1){\makebox(1,1){1}}
\put(14,1){\makebox(1,1){2}}
\put(15,1){\makebox(1,1){3}}
\put(15,0){\makebox(1,1){4}}
\put(16,0){\makebox(1,1){5}}
\put(17,1){\makebox(1,1){6}}
\multiput(13.5,0.5)(1,0){2}{\circle*{0.4}}
\multiput(16.5,1.5)(1,-1){2}{\circle*{0.4}}
\put(21,1){\makebox(1,1){4}}
\put(24,1.5){\line(1,0){7}}
\multiput(25,1.35)(1,0){6}{\line(0,1){0.3}}
\multiput(25,1.5)(1,0){2}{\circle*{0.4}}
\multiput(29,1.5)(1,0){2}{\circle*{0.4}}
\put(24.5,0){\makebox(1,1){1}}
\put(25.5,0){\makebox(1,1){2}}
\put(28.5,0){\makebox(1,1){5}}
\put(29.5,0){\makebox(1,1){6}}
\end{picture}

\begin{picture}(31,3)(1,0)
\linethickness{0.3pt}
\multiput(0,1)(0,1){2}{\line(1,0){5}}
\put(0,0){\line(1,0){1}}
\multiput(0,1)(1,0){6}{\line(0,1){1}}
\multiput(0,0)(1,0){2}{\line(0,1){1}}
\put(0,1){\makebox(1,1){1}}
\put(1,1){\makebox(1,1){2}}
\put(2,1){\makebox(1,1){3}}
\put(3,1){\makebox(1,1){4}}
\put(0,0){\makebox(1,1){5}}
\put(4,1){\makebox(1,1){6}}
\put(8,1){\makebox(1,1){2}}
\put(10.5,0.9){\makebox(1,1){$\rightarrow$}}
\multiput(13,1)(0,1){2}{\line(1,0){5}}
\put(16,0){\line(1,0){1}}
\multiput(13,1)(1,0){6}{\line(0,1){1}}
\multiput(16,0)(1,0){2}{\line(0,1){1}}
\put(13,1){\makebox(1,1){1}}
\put(14,1){\makebox(1,1){2}}
\put(15,1){\makebox(1,1){3}}
\put(16,1){\makebox(1,1){4}}
\put(16,0){\makebox(1,1){5}}
\put(17,1){\makebox(1,1){6}}
\multiput(13.5,0.5)(1,0){3}{\circle*{0.4}}
\put(17.5,0.5){\circle*{0.4}}
\put(21,1){\makebox(1,1){4}}
\put(24,1.5){\line(1,0){7}}
\multiput(25,1.35)(1,0){6}{\line(0,1){0.3}}
\multiput(25,1.5)(1,0){3}{\circle*{0.4}}
\put(30,1.5){\circle*{0.4}}
\put(24.5,0){\makebox(1,1){1}}
\put(25.5,0){\makebox(1,1){2}}
\put(26.5,0){\makebox(1,1){3}}
\put(29.5,0){\makebox(1,1){6}}
\end{picture}
\end{center}
  \caption{Examples of eigenstates of the SU(2) HSM 
    with $N=6$ sites in terms of spinons. The dots represent the spinons, the
    first tableau is the ground state.}
  \label{fig:foursitesu2}
\end{figure}
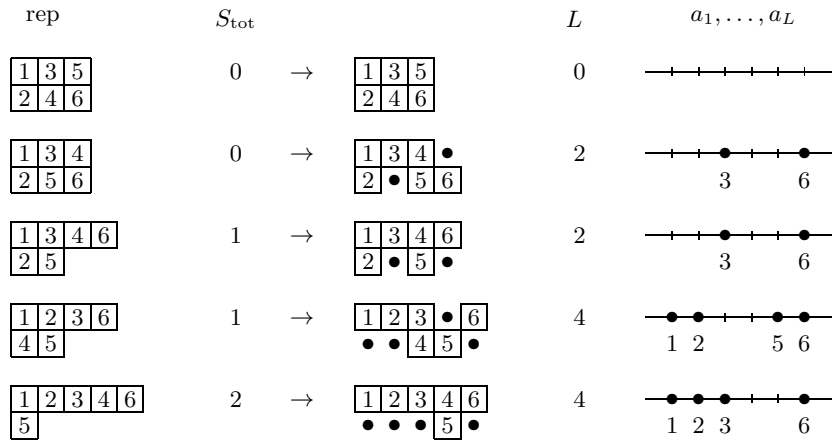
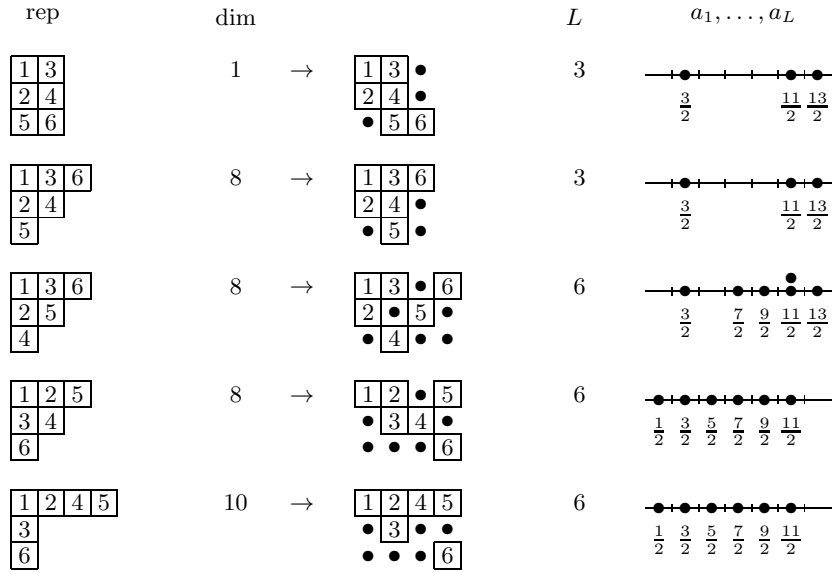
\begin{figure}[tb]
\setlength{\unitlength}{10pt}
\begin{center}
\begin{picture}(31,6)(1,0)
\linethickness{0.3pt}
\put(0.2,4){\makebox(2,1){rep}}
\put(7,4){\makebox(3,1){dim}}
\put(17.8,4){\makebox(7,1){$L$}}
\put(25.2,4){\makebox(5,1){$a_1,\ldots,a_L$}}
\multiput(0,0)(0,1){4}{\line(1,0){2}}
\multiput(0,0)(1,0){3}{\line(0,1){3}}
\put(0,2){\makebox(1,1){1}}
\put(0,1){\makebox(1,1){2}}
\put(1,2){\makebox(1,1){3}}
\put(1,1){\makebox(1,1){4}}
\put(0,0){\makebox(1,1){5}}
\put(1,0){\makebox(1,1){6}}
\put(8,2){\makebox(1,1){1}}
\put(10.5,1.9){\makebox(1,1){$\rightarrow$}}
\multiput(13,2)(0,1){2}{\line(1,0){2}}
\put(13,1){\line(1,0){3}}
\put(14,0){\line(1,0){2}}
\multiput(13,2)(1,0){3}{\line(0,1){1}}
\multiput(13,1)(1,0){3}{\line(0,1){1}}
\multiput(14,0)(1,0){3}{\line(0,1){1}}
\put(13,2){\makebox(1,1){1}}
\put(13,1){\makebox(1,1){2}}
\put(14,2){\makebox(1,1){3}}
\put(14,1){\makebox(1,1){4}}
\put(14,0){\makebox(1,1){5}}
\put(15,0){\makebox(1,1){6}}
\put(13.5,0.5){\circle*{0.4}}
\multiput(15.5,1.5)(0,1){2}{\circle*{0.4}}
\put(21,2){\makebox(1,1){3}}
\put(24,2.3){\line(1,0){7}}
\multiput(25,2.15)(1,0){6}{\line(0,1){0.3}}
\put(25.5,2.3){\circle*{0.4}}
\multiput(29.5,2.3)(1,0){2}{\circle*{0.4}}
\put(25,0.6){\makebox(1,1){$\frac{3}{2}$}}
\put(29,0.6){\makebox(1,1){$\frac{11}{2}$}}
\put(30,0.6){\makebox(1,1){$\frac{13}{2}$}}
\end{picture}

\begin{picture}(31,4)(1,0)
\linethickness{0.3pt}
\multiput(0,2)(0,1){2}{\line(1,0){3}}
\put(0,1){\line(1,0){2}}
\put(0,0){\line(1,0){1}}
\multiput(0,0)(1,0){2}{\line(0,1){3}}
\put(2,1){\line(0,1){2}}
\put(3,2){\line(0,1){1}}
\put(0,2){\makebox(1,1){1}}
\put(0,1){\makebox(1,1){2}}
\put(1,2){\makebox(1,1){3}}
\put(1,1){\makebox(1,1){4}}
\put(0,0){\makebox(1,1){5}}
\put(2,2){\makebox(1,1){6}}
\put(8,2){\makebox(1,1){8}}
\put(10.5,1.9){\makebox(1,1){$\rightarrow$}}
\multiput(13,2)(0,1){2}{\line(1,0){3}}
\put(13,1){\line(1,0){2}}
\put(14,0){\line(1,0){1}}
\multiput(13,2)(1,0){4}{\line(0,1){1}}
\multiput(13,1)(1,0){3}{\line(0,1){1}}
\multiput(14,0)(1,0){2}{\line(0,1){1}}
\put(13,2){\makebox(1,1){1}}
\put(13,1){\makebox(1,1){2}}
\put(14,2){\makebox(1,1){3}}
\put(14,1){\makebox(1,1){4}}
\put(14,0){\makebox(1,1){5}}
\put(15,2){\makebox(1,1){6}}
\multiput(13.5,0.5)(2,0){2}{\circle*{0.4}}
\put(15.5,1.5){\circle*{0.4}}
\put(21,2){\makebox(1,1){3}}
\put(24,2.3){\line(1,0){7}}
\multiput(25,2.15)(1,0){6}{\line(0,1){0.3}}
\put(25.5,2.3){\circle*{0.4}}
\multiput(29.5,2.3)(1,0){2}{\circle*{0.4}}
\put(25,0.6){\makebox(1,1){$\frac{3}{2}$}}
\put(29,0.6){\makebox(1,1){$\frac{11}{2}$}}
\put(30,0.6){\makebox(1,1){$\frac{13}{2}$}}
\end{picture}

\begin{picture}(31,4)(1,0)
\linethickness{0.3pt}
\multiput(0,2)(0,1){2}{\line(1,0){3}}
\put(0,1){\line(1,0){2}}
\put(0,0){\line(1,0){1}}
\multiput(0,0)(1,0){2}{\line(0,1){3}}
\put(2,1){\line(0,1){2}}
\put(3,2){\line(0,1){1}}
\put(0,2){\makebox(1,1){1}}
\put(0,1){\makebox(1,1){2}}
\put(1,2){\makebox(1,1){3}}
\put(0,0){\makebox(1,1){4}}
\put(1,1){\makebox(1,1){5}}
\put(2,2){\makebox(1,1){6}}
\put(8,2){\makebox(1,1){8}}
\put(10.5,1.9){\makebox(1,1){$\rightarrow$}}
\put(13,3){\line(1,0){2}}
\put(16,3){\line(1,0){1}}
\put(13,2){\line(1,0){4}}
\put(13,1){\line(1,0){3}}
\put(14,0){\line(1,0){1}}
\multiput(13,2)(1,0){5}{\line(0,1){1}}
\multiput(13,1)(1,0){4}{\line(0,1){1}}
\multiput(14,0)(1,0){2}{\line(0,1){1}}
\put(13,2){\makebox(1,1){1}}
\put(13,1){\makebox(1,1){2}}
\put(14,2){\makebox(1,1){3}}
\put(14,0){\makebox(1,1){4}}
\put(15,1){\makebox(1,1){5}}
\put(16,2){\makebox(1,1){6}}
\multiput(13.5,0.5)(1,1){2}{\circle*{0.4}}
\multiput(15.5,0.5)(0,2){2}{\circle*{0.4}}
\multiput(16.5,0.5)(0,1){2}{\circle*{0.4}}
\put(21,2){\makebox(1,1){6}}
\put(24,2.3){\line(1,0){7}}
\multiput(25,2.15)(1,0){6}{\line(0,1){0.3}}
\put(25.5,2.3){\circle*{0.4}}
\multiput(27.5,2.3)(1,0){4}{\circle*{0.4}}
\put(29.5,2.8){\circle*{0.4}}
\put(25,0.6){\makebox(1,1){$\frac{3}{2}$}}
\put(27,0.6){\makebox(1,1){$\frac{7}{2}$}}
\put(28,0.6){\makebox(1,1){$\frac{9}{2}$}}
\put(29,0.6){\makebox(1,1){$\frac{11}{2}$}}
\put(30,0.6){\makebox(1,1){$\frac{13}{2}$}}
\end{picture}

\begin{picture}(31,4)(1,0)
\linethickness{0.3pt}
\multiput(0,2)(0,1){2}{\line(1,0){3}}
\put(0,1){\line(1,0){2}}
\put(0,0){\line(1,0){1}}
\multiput(0,0)(1,0){2}{\line(0,1){3}}
\put(2,1){\line(0,1){2}}
\put(3,2){\line(0,1){1}}
\put(0,2){\makebox(1,1){1}}
\put(1,2){\makebox(1,1){2}}
\put(0,1){\makebox(1,1){3}}
\put(1,1){\makebox(1,1){4}}
\put(2,2){\makebox(1,1){5}}
\put(0,0){\makebox(1,1){6}}
\put(8,2){\makebox(1,1){8}}
\put(10.5,1.9){\makebox(1,1){$\rightarrow$}}
\put(13,3){\line(1,0){2}}
\put(16,3){\line(1,0){1}}
\put(13,2){\line(1,0){4}}
\put(14,1){\line(1,0){3}}
\put(16,0){\line(1,0){1}}
\multiput(13,2)(1,0){5}{\line(0,1){1}}
\multiput(14,1)(1,0){3}{\line(0,1){1}}
\multiput(16,0)(1,0){2}{\line(0,1){1}}
\put(13,2){\makebox(1,1){1}}
\put(14,2){\makebox(1,1){2}}
\put(14,1){\makebox(1,1){3}}
\put(15,1){\makebox(1,1){4}}
\put(16,2){\makebox(1,1){5}}
\put(16,0){\makebox(1,1){6}}
\multiput(13.5,0.5)(0,1){2}{\circle*{0.4}}
\multiput(14.5,0.5)(2,1){2}{\circle*{0.4}}
\multiput(15.5,0.5)(0,2){2}{\circle*{0.4}}
\put(21,2){\makebox(1,1){6}}
\put(24,2.3){\line(1,0){7}}
\multiput(25,2.15)(1,0){6}{\line(0,1){0.3}}
\multiput(24.5,2.3)(1,0){6}{\circle*{0.4}}
\put(24,0.6){\makebox(1,1){$\frac{1}{2}$}}
\put(25,0.6){\makebox(1,1){$\frac{3}{2}$}}
\put(26,0.6){\makebox(1,1){$\frac{5}{2}$}}
\put(27,0.6){\makebox(1,1){$\frac{7}{2}$}}
\put(28,0.6){\makebox(1,1){$\frac{9}{2}$}}
\put(29,0.6){\makebox(1,1){$\frac{11}{2}$}}
\end{picture}

\begin{picture}(31,4)(1,0)
\linethickness{0.3pt}
\multiput(0,2)(0,1){2}{\line(1,0){4}}
\multiput(0,0)(0,1){2}{\line(1,0){1}}
\multiput(0,0)(1,0){2}{\line(0,1){3}}
\multiput(2,2)(1,0){3}{\line(0,1){1}}
\put(0,2){\makebox(1,1){1}}
\put(1,2){\makebox(1,1){2}}
\put(0,1){\makebox(1,1){3}}
\put(2,2){\makebox(1,1){4}}
\put(3,2){\makebox(1,1){5}}
\put(0,0){\makebox(1,1){6}}
\put(8,2){\makebox(1,1){10}}
\put(10.5,1.9){\makebox(1,1){$\rightarrow$}}
\multiput(13,2)(0,1){2}{\line(1,0){4}}
\put(14,1){\line(1,0){1}}
\multiput(16,0)(0,1){2}{\line(1,0){1}}
\multiput(13,2)(1,0){5}{\line(0,1){1}}
\multiput(14,1)(1,0){2}{\line(0,1){1}}
\multiput(16,0)(1,0){2}{\line(0,1){1}}
\put(13,2){\makebox(1,1){1}}
\put(14,2){\makebox(1,1){2}}
\put(14,1){\makebox(1,1){3}}
\put(15,2){\makebox(1,1){4}}
\put(16,2){\makebox(1,1){5}}
\put(16,0){\makebox(1,1){6}}
\multiput(15.5,1.5)(1,0){2}{\circle*{0.4}}
\put(13.5,1.5){\circle*{0.4}}
\multiput(13.5,0.5)(1,0){3}{\circle*{0.4}}
\put(21,2){\makebox(1,1){6}}
\put(24,2.3){\line(1,0){7}}
\multiput(25,2.15)(1,0){6}{\line(0,1){0.3}}
\multiput(24.5,2.3)(1,0){6}{\circle*{0.4}}
\put(24,0.6){\makebox(1,1){$\frac{1}{2}$}}
\put(25,0.6){\makebox(1,1){$\frac{3}{2}$}}
\put(26,0.6){\makebox(1,1){$\frac{5}{2}$}}
\put(27,0.6){\makebox(1,1){$\frac{7}{2}$}}
\put(28,0.6){\makebox(1,1){$\frac{9}{2}$}}
\put(29,0.6){\makebox(1,1){$\frac{11}{2}$}}
\end{picture}
\end{center}
  \caption{Examples of eigenstates of the SU(3) HSM 
    with $N=6$ sites in terms of colorons.}
  \label{fig:sixsitesu3}
\end{figure}

Now, there is a one-to-one correspondence between these Young tableaux and the
non-interacting many-spinon states, \ie the eigenstates of the HSM.  The
principle is illustrated for a few representations of an SU(2) chain with six
sites in Fig.~\ref{fig:foursitesu2}, and for an SU(3) chain in
Fig.~\ref{fig:sixsitesu3}.  In each Young tableau we shift boxes to the right
such that each box is below or in the column to the right of the box with the
preceding number.  Each missing box in the resulting, extended tableaux
represents a spinon, to which we assign a spinon momentum number (SMN) $a_i$
as follows: For an SU(2) chain, it is simply given by the number in the box in
the same column.  For a general SU($n$) chain, the SMN's for the spinons in
each column are given by a sequence of numbers (half-integers for $n$ odd,
integers for $n$ even) with integer spacings such that the arithmetic mean
equals the arithmetic mean of the numbers in the boxes of the column, as
illustrated in the examples presented in Fig.~\ref{fig:sixsitesu3}.  The
extended tableaux provide us with the total SU($n$) spin of each multiplet
(given by the original Young tableau), as well as the number $L$ of spinons
present and the individual single-spinon momenta $p_1,\ldots,p_L$ given in
terms of the SMN's as
\begin{equation}
  \label{eq:singlespinonmom}
  p_i=\frac{2\pi}{N}\:\frac{a_i-\frac{1}{2}}{n},
\end{equation}
which implies $0\le p_i\le\frac{2\pi}{n}$ for $N\to\infty$.  The total
momentum and Haldane--Shastry energies of the many-spinon states are
\begin{equation}
  p=p_0+\sum_{i=1}^L p_i,\quad E=E_0+\sum_{i=1}^L \epsilon(p_i),
  \label{eq:Lspinonenergy}
\end{equation} 
where $p_0$ and $E_0$ denote the ground-state momentum and energy
given by 
\begin{equation}
  \label{eq:pzeroezero}
  p_0=-\frac{(n\!-\!1)\pi}{n}\:N,\ \
  E_0=-\frac{\pi^2}{12}\!\left(\frac{n\!-\!2}{n}N+\frac{2n\!-\!1}{N}\right)\!,
\end{equation}
and the single-spinon dispersion is
\begin{equation}
  \label{eq:spinonenergyepsilon}
  \epsilon(p)=\frac{n}{4}\,p\left(\frac{2\pi}{n}-p\right)
    +\frac{n^2-1}{12n}\,\frac{\pi^2}{N^2}.
\end{equation}
This formalism provides the complete spectrum of the HSM~\cite{GS07}. It
is easy to see that the momentum spacings for spin-polarised spinons predicted
by this formalism reproduce (\ref{eq:twospinonmomenta}) for general $n$, and
that spinons transform under the representation $\boldsymbol{\bar{n}}$ of
SU($n$).

\section{Yangians}\label{sec:yangians}

In this section we discuss the Yangian of sl$_n$.  Let $\{I^a\}$ be an
orthonormal basis with respect to a scalar product $\langle .\,,.\rangle$ of
sl$_n$. We will use
\begin{equation}
\langle X,Y\rangle =\mathrm{tr}(XY),\quad X,Y\in\mathrm{sl}_n.
\label{eq:yangianslnscp}
\end{equation}
For example, an orthonormal basis of sl$_2$ is then given by
$I^a=\sqrt{2}S^a=\sigma^a/\sqrt{2}$ with the Pauli matrices $\sigma^a,
a=1,2,3$.  The operators $I^a$ fulfil the algebra
\begin{equation}
\comm{I^a}{I^b}=c^{abc}I^c,\quad a,b,c=1,\ldots,n^2-1,
\label{eq:yangiansunalgebra}
\end{equation}
where $c^{abc}$ are the structure constants, \eg
$c^{abc}=\mathrm{i}\sqrt{2}\varepsilon^{abc}$ for sl$_2$.

The Yangian Y(sl$_n$)~\cite{Drinfeld85} of sl$_n$ is the
infinite-dimensional associative algebra over $\mathbb{C}$ generated
by the elements $I^a$, $\mathcal{I}^a$ with defining relations
\begin{eqnarray}
&&\hspace{-20mm}\comm{I^a}{I^b}=c^{abc}I^c,
\qquad\comm{I^a}{\mathcal{I}^b}=c^{abc}\,\mathcal{I}^c,
\label{eq:yangiandefeq1}\\
&&\hspace{-20mm}\Bigl[\mathcal{I}^a,\comm{\mathcal{I}^b}{I^c}\Bigr]-
\Bigl[I^a,\comm{\mathcal{I}^b}{\mathcal{I}^c}\Bigr]=
\mathfrak{c}^{abcdef}\left\{I^d,I^e,I^f\right\}\!,
\label{eq:yangiandefeq2}\\
&&\hspace{-20mm}
\Bigl[\comm{\mathcal{I}^a}{\mathcal{I}^b},\comm{I^i}{\mathcal{I}^j}\Bigr]+
\Bigl[\comm{\mathcal{I}^i}{\mathcal{I}^j},\comm{I^a}{\mathcal{I}^b}\Bigr]=
\left(\mathfrak{c}^{abcdef}c^{ijc}+\mathfrak{c}^{ijcdef}c^{abc}\right)
\left\{I^d,I^e,\mathcal{I}^f\right\}\!,
\label{eq:yangiandefeq3}
\end{eqnarray}
where repeated indices are summed over and
\begin{equation}
\hspace{-10mm}
\mathfrak{c}^{abcdef}=\frac{1}{24}\,c^{adi}c^{bej}c^{cfk}c^{ijk},\quad
\left\{X^a,X^b,X^c\right\}=\sum_{\pi\in S_3}
X^{\pi(a)}X^{\pi(b)}X^{\pi(c)}.
\end{equation}
Y(sl$_n$) has a Hopf algebra structure with comultiplication $\Delta:
\mathrm{Y(sl}_n)\rightarrow \mathrm{Y(sl}_n)\otimes \mathrm{Y(sl}_n)$ given by
\begin{equation}
\hspace{-10mm}
\Delta(I^a)= I^a\otimes 1 + 1\otimes I^a,\quad
\Delta(\mathcal{I}^a)=\mathcal{I}^a\otimes 1+1\otimes\mathcal{I}^a
-\frac{1}{2}\,c^{abc}I^b\otimes I^c.\label{eq:comultiplication}
\end{equation}
The counit $\epsilon: \mathrm{Y(sl}_n)\rightarrow\mathbb{C}$ and the
antipode $\mathcal{S}: \mathrm{Y(sl}_n)\rightarrow \mathrm{Y(sl}_n)$
will not be used in this article, there definitions can be found in
the literature~\cite{Drinfeld85,ChariPressley98}.  The defining
relations (\ref{eq:yangiandefeq1})--(\ref{eq:yangiandefeq3}) depend on
the choice of the scalar product (\ref{eq:yangianslnscp}) but, up to
isomorphism, the Hopf algebra Y(sl$_n$) does
not~\cite{ChariPressley98}. We have chosen (\ref{eq:yangianslnscp}) in
order to match the notations of~\cite{ChariPressley90,ChariPressley96}
for the representation theory of Y(sl$_n$).

The algebra generated by the total spin (\ref{eq:su3-Jsymmetry}) and
rapidity operator (\ref{eq:su3-Upsilon}) is recovered with the
identifications
\begin{equation}
S^a=\frac{1}{\sqrt{2}}\,I^a \mbox{ or } J^a=\frac{1}{\sqrt{2}}\,I^a,\quad 
\Lambda^a=\frac{1}{\sqrt{2}}\,\mathcal{I}^a,
\label{eq:yangianindentification}
\end{equation}
hence, the Yangian constitutes a symmetry of the
HSM~\cite{Haldane-92,Hikami95npb}.  The comultiplication defines the action of
the operators (\ref{eq:yangianindentification}) on two-particle states, and
being a homomorphism it is consistent on three-particle states.

There is another realization of the Yangian, first given by Drinfel'd in
1988~\cite{Drinfeld88}, which will be used to set the representation theory of
the Yangian in the next section.  It is based on the Cartan--Weyl
basis~\cite{Cornwell84vol2} 
of sl$_n$, which is for $n=2$ is given in terms of the spin operators by
\begin{equation}
H_1=2S^z,\quad X^\pm_1=S^\pm=S^x\pm\mathrm{i} S^y,
\end{equation}
whereas for $n=3$ we have explicitly
\begin{eqnarray}
&&H_1=2J^3,\quad H_2=\sqrt{3}J^8-J^3,\\
&&X^\pm_1=I^\pm=J^1\pm\mathrm{i} J^2,\quad
X^\pm_2=U^\pm=J^6\pm\mathrm{i} J^7,\quad
V^\pm=J^4\pm\mathrm{i} J^5.
\end{eqnarray}
The operators $I^\pm$ and $U^\pm$ are sufficient to span sl$_3$ as we can
re-express the non-simple root as $V^\pm=I^\pm+U^\pm$.

With the identifications $H_{i,0}=H_i$ and $X^\pm_{i,0}=X^\pm_i$ the
relations of the Yangian Y(sl$_n)$ read as
follows~\cite{Drinfeld88,ChariPressley98} ($1\le i\le n-1$,
$k\in\mathbb{N}_0$):
\begin{eqnarray}
&&\hspace{-20mm}
\comm{H_{i,k}^{\phantom{\pm}}}{H_{j,l}^{\phantom{\pm}}}=0,\quad
\comm{H_{i,0}^{\phantom{\pm}}}{X^\pm_{j,k}}=\pm B_{ij} X^\pm_{j,k},\quad
\comm{X^+_{i,k}}{X^-_{j,l}}=\delta_{ij}H_{i,k+l}^{\phantom{\pm}},\\
&&\hspace{-20mm}
\comm{H_{i,k+1}^{\phantom{\pm}}}{X^\pm_{j,l}}-
\comm{H^{\phantom{\pm}}_{i,k}}{X^\pm_{j,l+1}}=
\pm\frac{1}{2}B_{ij}\left(H_{i,k}^{\phantom{\pm}}X^\pm_{j,l}+
X^\pm_{j,l}H_{i,k}^{\phantom{\pm}}\right)\!,\\
&&\hspace{-20mm}
\comm{X^\pm_{i,k+1}}{X^\pm_{j,l}}-\comm{X^\pm_{i,k}}{X^\pm_{j,l+1}}=
\pm\frac{1}{2}B_{ij}
\left(X^\pm_{i,k}X^\pm_{j,l}+X^\pm_{j,l}X^\pm_{i,k}\right)\!,\\
&&\hspace{-20mm}
\Bigl[X^\pm_{i,k_1},\comm{X^\pm_{i,k_2}}{X^\pm_{j,l}}\Bigr]+
\Bigl[X^\pm_{i,k_2},\comm{X^\pm_{i,k_1}}{X^\pm_{j,l}}\Bigr]=0,
\qquad\mathrm{for}\quad i=\pm(j+1),\\
&&\hspace{-20mm}
\comm{X^\pm_{i,k}}{X^\pm_{j,l}}=0,
\qquad\mathrm{for}\quad i\neq j,\pm(j+1),
\end{eqnarray}
where $B_{ii}=2$, $B_{i,i+1}=B_{i,i-1}=-1$, and $B_{ij}=0$ otherwise.

For simplicity we state the isomorphism between the two realizations of
Y(sl$_n$) only for the diagonal operators in the cases $n=2$ and $n=3$. For
Y(sl$_2$) it is given by
\begin{equation}
S^z \mapsto \frac{1}{2} H_{1,0},\quad
\Lambda^z\mapsto \frac{1}{2} H_{1,1}+
\frac{1}{4}\left(S^+S^-+S^-S^+-H_{1,0}^2\right)\!,
\label{eq:su2isomorpihsm}
\end{equation}
while for Y(sl$_3$) it reads
\begin{eqnarray}
&&\hspace{-20mm}
J^3\mapsto \frac{1}{2} H_{1,0},\quad 
J^8\mapsto \frac{1}{2\sqrt{3}}H_{1,0}+\frac{1}{\sqrt{3}}H_{2,0},
\label{eq:cwbj8}\\
&&\hspace{-20mm}
\Lambda^3\mapsto \frac{1}{2} H_{1,1}-\frac{1}{4}H_{1,0}^2
+\frac{1}{4}\!\left(I^+I^-\!+\!I^-I^+\right)
-\frac{1}{8}\!\left(U^+U^-\!+\!U^-U^+\!-\!V^+V^-\!-\!V^-V^+\right)\!,
\label{eq:cwblambda3}\\
&&\hspace{-20mm}
\Lambda^8\mapsto \frac{1}{2\sqrt{3}}H_{1,1}+\frac{1}{\sqrt{3}}H_{2,1}
-\frac{1}{4\sqrt{3}}H_{1,0}^2-\frac{1}{2\sqrt{3}}H_{2,0}^2\nonumber\\
&&\hspace{-10mm}
+\frac{\sqrt{3}}{8}\left(U^+U^-+U^-U^++V^+V^-+V^-V^+\right)\!.
\label{eq:cwblambda8}
\end{eqnarray}
Here, we have used the short-hand notations $S^\pm=X_{1,0}^\pm$ as well as
$I^\pm=X_{1,0}^\pm$, $U^\pm=X_{2,0}^\pm$, and $V^\pm=X_{1,0}^\pm+X_{2,0}^\pm$.

\section{Representations of Yangians}\label{sec:repofyangians}

The representation theory of
Y(sl$_n$)~\cite{ChariPressley90,ChariPressley96,ChariPressley98,Molev02}
is based on the existence of the evaluation homomorphism, which
connects Y(sl$_n$) with the universal enveloping algebra U(sl$_n$)
of sl$_n$. For all $\xi\in\mathbb{C}$,
$\mathrm{ev_\xi}:\mathrm{Y(sl}_n)\to\mathrm{U(sl}_n)$ is given by
\begin{equation}
I^a\mapsto I^a,\quad
\mathcal{I}^a\mapsto\xi I^a+\frac{1}{4}\sum_{b,c=1}^{n^2-1}\mathrm{tr}
\Bigl(I^a\left(I^bI^c+I^cI^b\right)\Bigr)\,I^bI^c,
\label{eq:yangianevdef3}
\end{equation}
where the trace is computed by representing the $I^a$ as $n\times n$ matrices.
In general, given a representation of sl$_n$ one obtains a one-parameter
family of Y(sl$_n$) representations via the pull-back of the evaluation
homomorphism. Explicitly, if $\phi$ is a representation of sl$_n$ on $V$,
$\phi_\xi=\phi\circ\mathrm{ev}_\xi$ is a representation of Y(sl$_n$) on $V$,
which is called the evaluation representation with spectral parameter $\xi$.

A representation $V$ of the Yangian Y(sl$_n$) is said to be highest weight, if
there exists a vector $v\in V$ such that $V=\mathrm{Y(sl}_n)\,v$ and
\begin{equation}
X_{i,k}^+ v=0, \qquad H_{i,k}^{\phantom{+}}v = d_{i,k}^{\phantom{+}}v,
\label{eq:defhighetweight}
\end{equation}
with an array of complex numbers $(d_{i,k})$.  In this case, $v$ is called the
Yangian highest weight state (YHWS) of $V$ and $(d_{i,k})$ its highest weight.
As it was shown by Drinfel'd~\cite{Drinfeld88}, the irreducible highest weight
representation $V$ of Y(sl$_n)$ with highest weight $(d_{i,k})$ is finite
dimensional if and only if there exist monic polynomials
$P_i\in\mathbb{C}[u]$, $1\le i\le n-1$ such that
\begin{equation}
\frac{P_i(u+1)}{P_i(u)}=1+\sum_{k=0}^\infty \frac{d_{i,k}}{u^{k+1}},
\label{eq:defpolynomial}
\end{equation}
in the sense that the right-hand side is a Laurent expansion of the
left-hand side about $u=\infty$~\cite{ChariPressley96}.
Hence, the Drinfel'd polynomials $P_i(u)$ classify the
finite-dimensional irreducible representations of the Yangian. The
$i$th fundamental representation of Y(sl$_n)$ with spectral parameter
$\xi\in\mathbb{C}$ is defined as the irreducible highest weight
representation with Drinfel'd polynomials
\begin{equation}
  P_i(u)=u-\xi,\qquad P_j(u)=1\quad\mathrm{for}\; j\neq i.
\end{equation}
We will denote the fundamental representation of Y(sl$_2$) with Drinfel'd
polynomial $P(u)=u-\xi$ by $V\left(\boldsymbol{\frac{1}{2}},\xi\right)$. It
can be constructed explicitly as the pull-back of the sl$_2$ representation
$\boldsymbol{\frac{1}{2}}$ under ev$_\xi$.  For Y(sl$_3$) we denote by
$V(\boldsymbol{3},\xi)$ and $V(\boldsymbol{\bar{3}},\xi)$ the
three-dimensional representations with Drinfel'd polynomials
$P_1(u)=u-\xi,\;P_2(u)=1$ and $P_1(u)=1,\;P_2(u)=u-\xi$, respectively. If
$V(\boldsymbol{3},\xi)$ and $V(\boldsymbol{\bar{3}},\xi)$ are realised as
evaluation representations, we obtain an additional shift in the spectral
parameter due to the trace on the right-hand side of (\ref{eq:yangianevdef3}).
For example, $V(\boldsymbol{\bar{3}},\xi)$ is obtained as evaluation
representation of the sl$_3$ representation $\boldsymbol{\bar{3}}$ with
evaluation parameter $\xi+2/3$ (see \ref{sec:yangianappev}).

\subsection{Representation theory of $Y(sl_2)$}\label{sec:yangianrepsl2}

In the following we denote the evaluation representation of the
$(m+1)$-di\-men\-sional sl$_2$ representation $\boldsymbol{\frac{m}{2}}$ with
spectral parameter $\xi\in\mathbb{C}$ by
$V\left(\boldsymbol{\frac{m}{2}},\xi\right)$. The Drinfel'd polynomial of
$V\left(\boldsymbol{\frac{m}{2}},\xi\right)$ is given
by~\cite{ChariPressley90}
\begin{equation}
P(u)=\left(u-\xi+\frac{m-1}{2}\right)\left(u-\xi+\frac{m-3}{2}\right)\cdots
\left(u-\xi-\frac{m-1}{2}\right)\!.
\label{eq:yangpolyrep2}
\end{equation}

Now, let $V_1$ and $V_2$ be two representations of Y(sl$_2$). The action of
$X\in\mathrm{Y(sl}_2)$ on the tensor product $V_1\otimes V_2$ is given by
$\Delta(X)$, where $\Delta$ is the comultiplication
(\ref{eq:comultiplication}).  We stress that due to the last term of
$\Delta(\mathcal{I}^a)$ $V_1$ and $V_2$ are intertwined.  In particular, as
$\Delta$ is not commutative, $V_1\otimes V_2$ and $V_2\otimes V_1$ are not
isomorphic in general.  In all cases, however, if $v_1^+$ and $v_2^+$ are the
YHWS's of $V_1$ and $V_2$, respectively, the vector $v_1^+\otimes v_2^+$ will
be the YHWS of $V_1\otimes V_2$.  The action of Y(sl$_2$) on $r$-fold tensor
products is defined by repeated application of $\Delta$.

The central theorem on the tensor product
$V=V\left(\boldsymbol{\frac{m_1}{2}},\xi_1\right)\otimes
V\left(\boldsymbol{\frac{m_2}{2}},\xi_2\right)$ is due to Chari and
Pressley~\cite{ChariPressley90,Tarasov85}:
\begin{enumerate}
\item[(i)] If $|\xi_1-\xi_2|\neq\frac{m_1+m_2}{2}-p+1$ for all
  $p\in\mathbb{N}$ with $0<p\le\mathrm{min}(m_1,m_2)$, then $V$ is irreducible
  as Y(sl$_2$) representation.
\item[(ii)] If $\xi_2-\xi_1=\frac{m_1+m_2}{2}-p+1$ for some $p\in\mathbb{N}$
  with $0<p\le\mathrm{min}(m_1,m_2)$, then $V$ has a unique proper Y(sl$_2)$
  subrepresentation $W$ generated by the highest weight vector of $V$.
  Explicitly, we have
\begin{displaymath}
\textstyle
W\cong V\left(\boldsymbol{\frac{p-1}{2}},\xi_1+\frac{m_1-p+1}{2}\right)\otimes 
V\left(\boldsymbol{\frac{m_1+m_2-p+1}{2}},\xi_2-\frac{m_1-p+1}{2}\right)
\end{displaymath}
and, as representation of sl$_2$, $W\cong
\boldsymbol{\frac{m_1+m_2}{2}}\oplus\cdots\oplus
\boldsymbol{\frac{m_1+m_2-2p+2}{2}}$.
\end{enumerate}
The third case, $\xi_1-\xi_2=\frac{m_1+m_2}{2}-p+1$ for some $p\in\mathbb{N}$
with $0<p\le\mathrm{min}(m_1,m_2)$, was also discussed
in~\cite{ChariPressley90} but will not be used in the study of the HSM.

In order to illustrate the two different situations, we consider the simplest
non-trivial case $V=V\left(\boldsymbol{\frac{1}{2}},\xi_1\right)\otimes
V\left(\boldsymbol{\frac{1}{2}},\xi_2\right)$ for $\xi_1<\xi_2$.  Clearly,
regarded as representation of sl$_2$ $V$ it decomposes as
$V\cong\boldsymbol{1}\oplus\boldsymbol{0}$.  If $\xi_2-\xi_1\neq 1$, then $V$
is irreducible as Y(sl$_2$) representation.  One can always find an operator
in Y(sl$_2$) which yields a singlet state when acting on a triplet state and
vice versa.  If $\xi_2-\xi_1=1$, however, $V$ contains a proper Y(sl$_2$)
subrepresentation $W\cong\boldsymbol{1}$ generated by the YHWS.  In
particular, there exists no operator in Y(sl$_2$) which yields the singlet
state when acting on a triplet state. However, it is possible to obtain a
triplet state when acting on the spin singlet state with an appropriate
operator. Hence, $V$ is not a direct sum of irreducible Y(sl$_2$)
representations.

The highest weight of $V=V\left(\boldsymbol{\frac{m_1}{2}},\xi_1\right)\otimes
V\left(\boldsymbol{\frac{m_2}{2}},\xi_2\right)$ is obtained from its Drinfel'd
polynomial, which is in the irreducible case simply given by the product of
the original polynomials~\cite{ChariPressley90}.  In the reducible case the
highest weight is determined by the highest weight component of $V$ using
(\ref{eq:yangpolyrep2}).

\subsection{Representation theory of $Y(sl_3)$}\label{sec:yangianrepsl3}

The representation theory of Y(sl$_3$) is not known in the same detail as it
is for Y(sl$_2$).  Although there exist irreducibility criteria for tensor
products of evaluation representations of
Y(sl$_n$)~\cite{Molev02}, an explicit description of the
irreducible subrepresentation of such tensor products including its spectral
parameter is missing.  We will restrict ourselves here to tensor products of
two fundamental representations of Y(sl$_3$).  There are three different
situations~\cite{ChariPressley96}:
\begin{enumerate}
\item[(i)] $V=V(\boldsymbol{3},\xi_1)\otimes V(\boldsymbol{3},\xi_2)$ is
  reducible as Y(sl$_3$) representation if and only if $\xi_1-\xi_2=\pm 1$. If
  $\xi_2-\xi_1=1$, then $V$ has a proper Y(sl$_3$) subrepresentation $W$
  generated by the highest weight vector of $V$ and, as representation of
  sl$_3$, $W\cong\boldsymbol{6}$.
\item[(ii)] $V=V(\boldsymbol{3},\xi_1)\otimes V(\boldsymbol{\bar{3}},\xi_2)$
  (or $V=V(\boldsymbol{\bar{3}},\xi_1)\otimes V(\boldsymbol{3},\xi_2)$) is
  reducible as Y(sl$_3$) representation if and only if $\xi_1-\xi_2=\pm 3/2$.
  If $\xi_2-\xi_1=3/2$, then $V$ has a proper Y(sl$_3$) subrepresentation $W$
  generated by the highest weight vector of $V$ and, as representation of
  sl$_3$, $W\cong\boldsymbol{8}$.
\item[(iii)] $V=V(\boldsymbol{\bar{3}},\xi_1)\otimes
  V(\boldsymbol{\bar{3}},\xi_2)$ is reducible as Y(sl$_3$) representation
  if and only if $\xi_1-\xi_2=\pm1$. If $\xi_1-\xi_2=1$, then $V$ has a proper
  Y(sl$_3$) subrepresentation $W$ not containing the highest weight vector of
  $V$ and, as representation of sl$_3$, $W\cong\boldsymbol{3}$.  If
  $\xi_2-\xi_1=1$, then $V$ has a proper Y(sl$_3$) subrepresentation $W$
  generated by the highest weight vector of $V$ and, as representation of
  sl$_3$, $W\cong\boldsymbol{\bar{6}}$.
\end{enumerate}
 
If the tensor product $V=V_1\otimes V_2$ is irreducible, the Drinfel'd
polynomials of $V$ are given by the products of the polynomials of $V_1$ and
$V_2$~\cite{ChariPressley98}. Furthermore, we show in \ref{sec:yangianappsc}
that the proper Y(sl$_3$) subrepresentation $W$ of
$V(\boldsymbol{\bar{3}},\xi)\otimes V(\boldsymbol{\bar{3}},\xi-1)$ is given by
$W\cong V\left(\boldsymbol{3},\xi-\frac{1}{2}\right)$.

\section{Spinons and representations of Y(sl$_{\boldsymbol{2}}$)}
\label{sec:yangiansu2}
\setlength{\unitlength}{10pt}

As mentioned above it is well-known~\cite{Haldane-92,Hikami95npb} that the
SU(2) HSM possesses the Yangian symmetry Y(sl$_2$) and therefore that its
Hilbert space decomposes into irreducible representations of Y(sl$_2$).  It is
also known~\cite{GS07} how to built up the Hilbert space of the HSM by
non-interacting spinon excitations.  In this section we will study the
relation between these many-spinon states and representations of Y(sl$_2$).

\subsection{One-spinon states}

We first derive the relation between the one-spinon states and the
fundamental representations of Y(sl$_2$).  Consider a chain with an
odd number of sites $N$. The individual one-spinon momenta are given
by~\cite{Haldane91prl1,Bernevig-01prb}
\begin{equation}
p = \frac{\pi}{2}-\frac{2\pi}{N}\left(\mu+\frac{1}{4}\right)\!,
\quad 0\le \mu\le\frac{N-1}{2},
\label{eq:yangianreponespinonmomentum}
\end{equation}
where we have assumed $N-1$ to be divisible by four, and thereby
restricted the momentum to $-\pi/2\le p\le\pi/2$.  The up-spin spinons are
YHWS's (they are annihilated by $S^+,\Lambda^+,\ldots$), their
spin currents are given by~\cite{Bernevig-01prb}
\begin{equation}
\Lambda^z\ket{\up}=\left(\frac{N-1}{4}-\mu\right)\ket{\up}\!.
\label{eq:yangiansc1}
\end{equation}
Here $\ket{\up}$ denotes the state with one up-spin spinon.

On the other hand the one-spinon states are represented by tableaux of
the form
\begin{center}
\begin{picture}(25,3)(0,0.5)
\linethickness{0.3pt}
\multiput(0,1)(1,0){10}{\line(0,1){2}}
\multiput(0,2)(0,1){2}{\line(1,0){9}}
\put(0,1){\line(1,0){3}}
\put(4,1){\line(1,0){5}}
\put(3.5,1.5){\circle*{0.4}}
\put(3,0){\makebox(1,1){$a$}}
\put(12,1.6){$\displaystyle a=N-2\mu,\quad 0\le \mu\le\frac{N-1}{2}$,}
\end{picture}
\end{center}
where we omit the superfluous numbers in the boxes of the tableaux. Note that
(\ref{eq:yangianreponespinonmomentum}) is recovered using
(\ref{eq:singlespinonmom})--(\ref{eq:pzeroezero}).  Now, 
(\ref{eq:su2isomorpihsm}) together with (\ref{eq:yangiansc1}) yields 
\begin{equation}
\hspace{-10mm}
H_{1,1}\ket{\up}=\left(2\Lambda^z-
\frac{1}{2}\Bigl[S^+S^-+S^-S^+-4(S^z)^2\Bigr]\right)\ket{\up}
=\left(a-\frac{N+1}{2}\right)\ket{\up}\!,
\end{equation}
where the term in squared brackets vanishes as the spinon possesses
spin $S=1/2$.

Hence, individual spinons transform under the Y(sl$_2$)
representation $V\left(\boldsymbol{\frac{1}{2}},\xi\right)$, where the
spectral parameter $\xi$ is in terms of the SMN given by
\begin{equation}
\xi=a-\frac{N+1}{2},\quad -\frac{N-1}{2}\le\xi\le\frac{N-1}{2}.
\label{eq:yangianrepxiasu2}
\end{equation}

\subsection{Two-spinon states}\label{sec:yangiantwospinons}

Let us consider two spin-polarised spinons represented by the tableau
\begin{center}
\begin{picture}(25,4.5)(0,-0.5)
\linethickness{0.3pt}
\multiput(0,1)(1,0){10}{\line(0,1){2}}
\multiput(0,2)(0,1){2}{\line(1,0){9}}
\put(0,1){\line(1,0){3}}
\put(4,1){\line(1,0){2}}
\put(7,1){\line(1,0){2}}
\multiput(3.5,1.5)(3,0){2}{\circle*{0.4}}
\put(3.1,-0.2){\makebox(1,1){$a_1$}}
\put(4.6,-0.2){\makebox(1,1){$<$}}
\put(6.1,-0.2){\makebox(1,1){$a_2$}}
\put(12,2.7){$a_1=N-2\mu-1,\quad a_2=N-2\nu,$}
\put(12,0){$\displaystyle 0\le\nu\le\mu\le\frac{N-2}{2}$.}
\end{picture}
\end{center}
Note that, as $N$ is even, $a_1$ is always odd and $a_2$ is always even. There
are two fundamentally different cases. If $a_2-a_1>1$, there exists a
two-spinon singlet with the same SMN's (and hence the same energy), whereas
for $a_2-a_1=1$ there exists no corresponding singlet. Graphically we have 
\begin{center}
\begin{picture}(28,3.5)
\linethickness{0.3pt}
\multiput(6,1)(1,0){10}{\line(0,1){2}}
\multiput(6,2)(0,1){2}{\line(1,0){9}}
\put(6,1){\line(1,0){3}}
\put(10,1){\line(1,0){2}}
\put(13,1){\line(1,0){2}}
\multiput(9.5,1.5)(3,0){2}{\circle*{0.4}}
\put(0,1.5){\makebox(3,1){$a_2-a_1>1:$}}
\put(16,1.5){\makebox(2,1){and}}

\multiput(19,1)(1,0){10}{\line(0,1){2}}
\put(19,3){\line(1,0){6}}
\put(26,3){\line(1,0){2}}
\put(19,2){\line(1,0){9}}
\put(19,1){\line(1,0){3}}
\put(23,1){\line(1,0){5}}
\multiput(22.5,1.5)(3,1){2}{\circle*{0.4}}
\end{picture}
\begin{picture}(28,3)(0,0.5)
\linethickness{0.3pt}
\multiput(6,1)(1,0){4}{\line(0,1){2}}
\multiput(11,1)(1,0){5}{\line(0,1){2}}
\put(10,2){\line(0,1){1}}
\multiput(6,2)(0,1){2}{\line(1,0){9}}
\put(6,1){\line(1,0){3}}
\put(11,1){\line(1,0){4}}
\multiput(9.5,1.5)(1,0){2}{\circle*{0.4}}
\put(0,1.5){\makebox(3,1){$a_2-a_1=1:$}}
\put(15.8,1.5){\makebox(3,1){only.}}
\end{picture}
\end{center}

This can be understood by applying the representation theory of Y(sl$_2$). In
both cases the two spinons transform under the product representation $V=
V\left(\boldsymbol{\frac{1}{2}},\xi_1\right)\otimes
V\left(\boldsymbol{\frac{1}{2}},\xi_2\right)$, where the spectral parameters
are given by $\xi_{1,2}=a_{1,2}-(N+1)/2$, respectively. In the first case we
have $\xi_2-\xi_1>1$, hence by Sec.~\ref{sec:yangianrepsl2}.i $V$ is irreducible
as Y(sl$_2$) representation.  As sl$_2$ representation we have
$V\cong\boldsymbol{1}\oplus\boldsymbol{0}$, \ie the triplet and singlet
represented by the two tableaux shown above.  $V$ is generated by its YHWS,
which is the spin-polarised two-spinon state
$\ket{\up\up}=\ket{\up}\otimes\ket{\up}$.  In particular, the operator
$\Lambda^zS^-\in\mathrm{Y(sl}_2)$ yields the two-spinon singlet state when
acting on the YHWS,
$\Lambda^zS^-\ket{\up\up}\propto\ket{\up\dw}-\ket{\dw\up}$, while leaving the
individual spinon momenta and hence the energy unchanged.

In the second case we have $\xi_2-\xi_1=1$, and by Sec.~\ref{sec:yangianrepsl2}.ii
$V$ is reducible.  The YHWS of $V$, which is again $\ket{\up\up}$, generates
the proper Y(sl$_2$) subrepresentation $W\cong\boldsymbol{1}$, \ie only the
triplet states are generated by $\ket{\up\up}$ and, in particular, it is
$\Lambda^zS^-\ket{\up\up}=0$.  This is reflected by the existence of only one
tableau if the SMN's fulfil $a_2-a_1=1$, and is consistent with results drawn
from the asymptotic Bethe Ansatz for the HSM~\cite{Haldane-92,Bernard-93} as
well as conformal field theory spectra~\cite{Bernard-94}.

The loss of the singlet, \ie its non-existence in the spinon Hilbert space, is
a manifestation of the fractional statistics of the spinons.  The momentum
spacings for two spinons with individual momenta $p_1$ and $p_2$ ($p_2>p_1$)
are in general given by $p_2-p_1=2\pi(1/2+\ell)/N$,
$\ell\in\mathbb{N}_0$~\cite{GS05prb}.  When the spinons occupy adjacent
momenta, $p_2-p_1=\pi/N$, only the triplet exists, which is mathematically
implemented by the requirement of irreducibility under Y(sl$_2$)
transformations.  In analogy, the Pauli principle enforces two electrons with
identical momenta to form a spin singlet, whereas otherwise a spin triplet
exists as well.  The difference between electrons and spinons is, however,
that the wave function of free electrons factorises in a spin part,
transforming under SU(2), as well as a momentum part, transforming under
(lattice) translations; the product of both has to be antisymmetric under
permutations.  In contrast we cannot factorise spin and momentum of the
spinons, as both are incorporated in the representations of Y(sl$_2$) (the
lattice translations are implemented as shifts of the spectral parameter
$\xi$). It seems that this entanglement of spin and momentum makes it
impossible to implement the fractional statistics by the requirement of a
definite transformation law under permutations of the spinons. In fact, this
requirement is replaced by the postulate of irreducibility under Yangian
transformations.

The spin current of the polarised two-spinon state $\ket{\up\up}$ is easily
obtained from Drinfel'd polynomial of the irreducible subrepresentation of
$V\left(\boldsymbol{\frac{1}{2}},\xi_1\right)\otimes
V\left(\boldsymbol{\frac{1}{2}},\xi_2\right)$, which is given by
$P(u)=(u-\xi_1)(u-\xi_2)$. Hence, with (\ref{eq:defhighetweight})
and (\ref{eq:defpolynomial}) we find
\begin{equation}
H_{1,1}\ket{\up\up}=(\xi_1+\xi_2+1)\ket{\up\up}=(N-2\mu-2\nu-1)\ket{\up\up}\!,
\end{equation}
and with (\ref{eq:su2isomorpihsm}) we obtain the physical spin current
\begin{equation}
\Lambda^z\ket{\up\up}=\left(\frac{N-2}{2}-\mu-\nu\right)\ket{\up\up}\!,
\end{equation}
which equals the result obtained using explicit wave
functions~\cite{Bernevig-01prb}.

\subsection{Many-spinon states}\label{sec:yangianmanyspinons}

If three spinons are present, there are three different cases, which are
graphically represented by
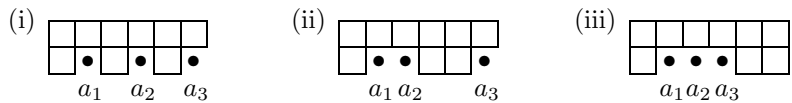
\begin{center}
\begin{picture}(28,4)
\linethickness{0.3pt}
\multiput(0,1)(1,0){6}{\line(0,1){2}}
\put(6,2){\line(0,1){1}}
\multiput(0,2)(0,1){2}{\line(1,0){6}}
\multiput(0,1)(2,0){3}{\line(1,0){1}}
\multiput(1.5,1.5)(2,0){3}{\circle*{0.4}}
\put(1.1,-0.2){\makebox(1,1){$a_1$}}
\put(3.1,-0.2){\makebox(1,1){$a_2$}}
\put(5.1,-0.2){\makebox(1,1){$a_3$}}
\put(-1.5,2.5){\makebox(1,1){(i)}}

\multiput(11,1)(1,0){2}{\line(0,1){2}}
\multiput(14,1)(1,0){3}{\line(0,1){2}}
\multiput(13,2)(4,0){2}{\line(0,1){1}}
\multiput(11,2)(0,1){2}{\line(1,0){6}}
\put(11,1){\line(1,0){1}}
\put(14,1){\line(1,0){2}}
\multiput(12.5,1.5)(1,0){2}{\circle*{0.4}}
\put(16.5,1.5){\circle*{0.4}}
\put(12.1,-0.2){\makebox(1,1){$a_1$}}
\put(13.2,-0.2){\makebox(1,1){$a_2$}}
\put(16.1,-0.2){\makebox(1,1){$a_3$}}
\put(9.35,2.5){\makebox(1,1){(ii)}}

\multiput(22,1)(1,0){2}{\line(0,1){2}}
\multiput(26,1)(1,0){3}{\line(0,1){2}}
\multiput(24,2)(1,0){2}{\line(0,1){1}}
\multiput(22,2)(0,1){2}{\line(1,0){6}}
\put(22,1){\line(1,0){1}}
\put(26,1){\line(1,0){2}}
\multiput(23.5,1.5)(1,0){3}{\circle*{0.4}}
\put(23.1,-0.2){\makebox(1,1){$a_1$}}
\put(24.1,-0.2){\makebox(1,1){$a_2$}}
\put(25.2,-0.2){\makebox(1,1){$a_3$}}
\put(20.2,2.5){\makebox(1,1){(iii)}}
\end{picture}
\end{center}
In all three cases we have to determine the irreducible subrepresentation of
$V=V\left(\boldsymbol{\frac{1}{2}},\xi_1\right)\linebreak \otimes
V\left(\boldsymbol{\frac{1}{2}},\xi_2\right)\otimes
V\left(\boldsymbol{\frac{1}{2}},\xi_3\right)$, where the spectral parameters
are given by $\xi_i=a_i-(N+1)/2$.  In the first case $V$ is
irreducible and generated by its YHWS $\ket{\up\up\up}$. As sl$_2$
representation we find $\boldsymbol{\frac{3}{2}}\oplus\boldsymbol{\frac{1}{2}}
\oplus\boldsymbol{\frac{1}{2}}$, which is the complete eight-dimensional space
$\boldsymbol{\frac{1}{2}}\otimes\boldsymbol{\frac{1}{2}}
\otimes\boldsymbol{\frac{1}{2}}$. The $\boldsymbol{\frac{3}{2}}$ is given by
the tableau (i) above; the tableaux representing the two
$\boldsymbol{\frac{1}{2}}$'s are obtained from (i) by moving either the second
or the third spinon to the first row.

In the second case we have $\xi_2-\xi_1=1$ and deduce using
Sec.~\ref{sec:yangianrepsl2}.ii that the irreducible
subrepresentation of
$V\left(\boldsymbol{\frac{1}{2}},\xi_1\right)\otimes
V\left(\boldsymbol{\frac{1}{2}},\xi_2\right)$ is
$V\left(\boldsymbol{1},\xi_1+\frac{1}{2}\right)$. The remaining tensor
product $V\left(\boldsymbol{1},\xi_1+\frac{1}{2}\right)\otimes
V\left(\boldsymbol{\frac{1}{2}},\xi_3\right)$ is irreducible, and as
sl$_2$ representation we obtain
$\boldsymbol{\frac{3}{2}}\oplus\boldsymbol{\frac{1}{2}}$, which is
only six-dimensional.  The loss of one $\boldsymbol{\frac{1}{2}}$ is
reflected by the fact that the second spinon in the tableau (ii) is
fixed to the lower row.  Note that this result is not affected if
$a_2$ and $a_3$ were adjacent instead of $a_1$ and $a_2$, although the
specific values of the spectral parameters will change.

In the third case the irreducible subrepresentation of
$V\left(\boldsymbol{\frac{1}{2}},\xi_1\right)\otimes
V\left(\boldsymbol{\frac{1}{2}},\xi_2\right)$ is again given by
$V\left(\boldsymbol{1},\xi_1+\frac{1}{2}\right)$, however, this time the
remaining tensor product is reducible as well; and its irreducible
subrepresentation is given by
$V\left(\boldsymbol{\frac{3}{2}},\xi_1+1\right)$. As sl$_2$ representation we
only have $\boldsymbol{\frac{3}{2}}$ which is represented by the tableau
(iii).

To give a more general examples let us first consider a six-site chain and the
four-spinon tableau
\begin{center}
\begin{picture}(6,2)(0,1)
\linethickness{0.3pt}
\multiput(0,2)(1,0){6}{\line(0,1){1}}
\multiput(2,1)(1,0){2}{\line(0,1){1}}
\multiput(0,2)(0,1){2}{\line(1,0){5}}
\put(2,1){\line(1,0){1}}
\multiput(0.5,1.5)(1,0){2}{\circle*{0.4}}
\multiput(3.5,1.5)(1,0){2}{\circle*{0.4}}
\end{picture}
\end{center}
The spin-polarised state in this multiplet, $\ket{\up\up\up\up}$, generates
the irreducible subrepresentation of
\begin{equation}
\textstyle
V\left(\boldsymbol{\frac{1}{2}},-\frac{5}{2}\right)\otimes 
V\left(\boldsymbol{\frac{1}{2}},-\frac{3}{2}\right)
\otimes V\left(\boldsymbol{\frac{1}{2}},\frac{3}{2}\right)\otimes 
V\left(\boldsymbol{\frac{1}{2}},\frac{5}{2}\right),
\end{equation}
which is given by $V(\boldsymbol{1},-2)\otimes V(\boldsymbol{1},2)$.  As
sl$_2$ representation this is given by
$\boldsymbol{2}\oplus\boldsymbol{1}\oplus\boldsymbol{0}$, which is represented
by the tableaux
\begin{center}
\begin{picture}(25,2.5)(0,1.2)
\linethickness{0.3pt}
\multiput(0,2)(1,0){6}{\line(0,1){1}}
\multiput(2,1)(1,0){2}{\line(0,1){1}}
\multiput(0,2)(0,1){2}{\line(1,0){5}}
\put(2,1){\line(1,0){1}}
\multiput(0.5,1.5)(1,0){2}{\circle*{0.4}}
\multiput(3.5,1.5)(1,0){2}{\circle*{0.4}}

\multiput(9,2)(1,0){6}{\line(0,1){1}}
\multiput(11,1)(1,0){3}{\line(0,1){1}}
\put(9,2){\line(1,0){5}}
\put(9,3){\line(1,0){3}}
\put(13,3){\line(1,0){1}}
\put(11,1){\line(1,0){2}}
\multiput(9.5,1.5)(1,0){2}{\circle*{0.4}}
\multiput(12.5,2.5)(1,-1){2}{\circle*{0.4}}

\multiput(18,2)(1,0){4}{\line(0,1){1}}
\multiput(20,1)(1,0){4}{\line(0,1){1}}
\put(18,2){\line(1,0){5}}
\multiput(18,3)(2,-2){2}{\line(1,0){3}}
\multiput(18.5,1.5)(1,0){2}{\circle*{0.4}}
\multiput(21.5,2.5)(1,0){2}{\circle*{0.4}}
\end{picture}
\end{center}

In the same way we can analyse the space generated by the YHWS of the 
seven-spinon tableau
\begin{center}
\begin{picture}(15,2)(0,1.2)
\linethickness{0.3pt}
\multiput(0,2)(1,0){13}{\line(0,1){1}}
\multiput(0,2)(0,1){2}{\line(1,0){12}}
\multiput(0,1)(1,0){2}{\line(0,1){1}}
\multiput(3,1)(1,0){2}{\line(0,1){1}}
\multiput(7,1)(1,0){3}{\line(0,1){1}}
\multiput(10,1)(1,0){2}{\line(0,1){1}}
\put(0,1){\line(1,0){1}}
\put(3,1){\line(1,0){1}}
\put(7,1){\line(1,0){2}}
\put(10,1){\line(1,0){1}}
\multiput(1.5,1.5)(1,0){2}{\circle*{0.4}}
\multiput(4.5,1.5)(1,0){3}{\circle*{0.4}}
\multiput(9.5,1.5)(2,0){2}{\circle*{0.4}}
\end{picture}
\end{center}
where $N=17$. We couple adjacent spinons according to
Sec.~\ref{sec:yangianrepsl2}.ii, and find the irreducible subrepresentation
to be
\begin{equation}
\textstyle
V\left(\boldsymbol{1},-\frac{11}{2}\right)\otimes 
V\left(\boldsymbol{\frac{3}{2}},-1\right)\otimes 
V\left(\boldsymbol{\frac{1}{2}},5\right)\otimes 
V\left(\boldsymbol{\frac{1}{2}},8\right),
\end{equation}
which as sl$_2$ representation reads
\begin{equation}
\textstyle
  \boldsymbol{1}\otimes\boldsymbol{\frac{3}{2}}\otimes
  \boldsymbol{\frac{1}{2}}\otimes\boldsymbol{\frac{1}{2}}
  \;=\;
  \boldsymbol{\frac{7}{2}}\oplus\boldsymbol{\frac{5}{2}}
  \oplus\boldsymbol{\frac{5}{2}}
  \oplus\boldsymbol{\frac{5}{2}}\oplus\boldsymbol{\frac{3}{2}}
  \oplus\boldsymbol{\frac{3}{2}}
  \oplus\boldsymbol{\frac{3}{2}}\oplus\boldsymbol{\frac{3}{2}}
  \oplus\boldsymbol{\frac{1}{2}}
  \oplus\boldsymbol{\frac{1}{2}}\oplus\boldsymbol{\frac{1}{2}}.
\end{equation}
The corresponding tableaux are easily constructed using that the
first, second, and fifth spinon are fixed to the lower row, and the
fourth spinon can only move to the upper row if the third one does.

The general scheme works as follows.  Any spin-polarised $m$-spinon state is
represented by a tableau with all spinons in the second row. The individual
spinon momenta are given in terms of the SMN's $a_i$.  The space generated by
this YHWS under the action of Y(sl$_2$) is the irreducible subrepresentation
$W$ of the tensor product
\begin{equation}
\textstyle
V=\,\bigotimes_{i=1}^m V\left(\boldsymbol{\frac{1}{2}},\xi_i\right),\quad 
\displaystyle\xi_i=a_i-\frac{N+1}{2}, 
\label{eq:yangiangeneralsu2product}
\end{equation}
where the $\xi_i$'s have ascending order. In order to construct $W$ one first
determines the irreducible subrepresentations of all partial products in
(\ref{eq:yangiangeneralsu2product}) which have consecutive spectral parameters
$\xi_{i+1}-\xi_i=1$ using Sec.~\ref{sec:yangianrepsl2}.ii. (Note that we can
without loss of generality begin with these products as the comultiplication
(\ref{eq:comultiplication}) is associative.)  The remaining tensor product is
then irreducible by repeated application of Sec.~\ref{sec:yangianrepsl2}.i
(for a proof see Ref.~\onlinecite{ChariPressley90}). The sl$_2$ contents is
determined by straightforward calculation.

To sum up, spinons in the HSM transform under the Y(sl$_2$) representation
$V\left(\boldsymbol{\frac{1}{2}},\xi\right)$, where the spectral parameter
$\xi$ is via (\ref{eq:yangianrepxiasu2}) and (\ref{eq:singlespinonmom})
directly connected to the spinon momentum.  All $m$-spinon states with given
individual momenta $p_1,\dots,p_m$ are generated by the YHWS of
(\ref{eq:yangiangeneralsu2product}), meaning that they span the irreducible
subrepresentation $W$. The complete Hilbert space is the direct sum of these
subspaces.  From a mathematical point of view the tableau
formalism~\cite{GS07} hence provides an algorithm to determine the sl$_2$
content of the irreducible subrepresentation of a tensor product of
fundamental Y(sl$_2$) representations (\ref{eq:yangiangeneralsu2product}) with
increasing spectral parameters (the restriction to integer or half-integer
spectral parameters $\xi_i$ is no limitation, since all $\xi_i$'s can be
shifted simultaneously and tensor products where the spacings $\xi_j-\xi_i$
are not integers are irreducible~\cite{ChariPressley90}).

\section{Colorons and representations of Y(sl$_{\boldsymbol{3}}$)}
\label{sec:yangiansu3}
\setlength{\unitlength}{10pt}

In this section we will investigate the relation between the Y(sl$_3$)
symmetry of the SU(3) HSM and its coloron excitations.

\subsection{One-coloron states}

Consider a chain with $N=3M-1$, $M\in\mathbb{N}$, sites. Then the one-coloron
momenta are given by~\cite{SG05epl}
\begin{equation}
p=\frac{4\pi}{3} -\frac{2\pi}{N}\left(\mu+\frac{1}{3}\right)\!,
\quad 0\le \mu\le (N-2)/3.
\label{eq:yangiancoloronmomentum}
\end{equation}
The SU(3) spin (or colour) currents of a yellow coloron $\ket{\y}$, which is a
Yangian lowest weight state, are
\begin{equation}
\frac{1}{\sqrt{3}}\Lambda^3\ket{\y}=
\Lambda^8\ket{\y}=-\frac{\sqrt{3}}{2}\left(\frac{N-2}{6}-\mu\right)\ket{\y}\!.
\end{equation}
In order to apply the representation theory of Y(sl$_3$) it will be
appropriate to work with YHWS's, that is magenta colorons $\ket{\m}$, instead.
As the fundamental representation $V(\boldsymbol{\bar{3}},\xi)$ of Y(sl$_3$)
can be explicitly realized as evaluation representation (see
\ref{sec:yangianappev}), we obtain the spin currents of $\ket{\m}$ to be
\begin{equation}
\Lambda^3\ket{\m}=0,\quad 
\Lambda^8\ket{\m}=\sqrt{3}\left(\frac{N-2}{6}-\mu\right)\ket{\m}\!.
\end{equation}

On the other hand a single coloron is represented by the tableau
\begin{center}
\begin{picture}(25,3.5)(0,1)
\linethickness{0.3pt}
\multiput(0,1)(1,0){10}{\line(0,1){3}}
\multiput(0,2)(0,1){3}{\line(1,0){9}}
\put(0,1){\line(1,0){3}}
\put(4,1){\line(1,0){5}}
\put(3.5,1.5){\circle*{0.4}}
\put(3,0){\makebox(1,1){$a$}}
\put(12,2){$\displaystyle a=N-3\mu-\frac{1}{2},\quad 0\le\mu\le\frac{N-2}{3}$.}
\end{picture}
\end{center}
The spectral parameter of $V(\boldsymbol{\bar{3}},\xi)$ is determined from the
eigenvalue of $H_{2,1}$ when acting on the YHWS $\ket{\m}$. Using
(\ref{eq:cwblambda8}) together with $H_{1,1}\ket{\m}=0$ we find
$H_{2,1}\ket{\m}=(\sqrt{3}\Lambda^8-1/4)\ket{\m}=(a-(2N+3)/4)\ket{\m}$.
Hence, colorons transform under the Y(sl$_3$) representation
$V(\boldsymbol{\bar{3}},\xi)$ with spectral parameter 
\begin{equation}
\xi=a-\frac{2N+3}{4},\quad -\frac{2N-3}{4}\le\xi\le\frac{2N-5}{4}.
\label{eq:yangianxidefinitionsu3}
\end{equation}
Note that although the allowed values for $\xi$ are not symmetrically
distributed around zero, the eigenvalues of the physical spin current
$\Lambda^8$ are.

\subsection{Two-coloron states}\label{sec:yangiantwocolorons}

Compared to the many-spinon states discussed above, the effect of the
fractional statistics on many-coloron states is rather complicated. We will
discuss in this and the next sections how the requirement of irreducibility
under Y(sl$_3$) transformations yields several restrictions on the allowed
SU(3) representations for many-coloron states.

Let us first consider two colorons with identical colours like
$\ket{\m\m}=\ket{\m}\otimes\ket{\m}$.  The individual coloron momenta
$p_1$ and $p_2$ with $p_2>p_1$ are spaced according to
$p_2-p_1=2\pi(2/3+\ell)/N$, $\ell\in\mathbb{N}_0$ (see
(\ref{eq:twospinonmomenta})).  Furthermore, for all
pairs of momenta satisfying this condition the SU(3) spin takes the
values $\boldsymbol{\bar{3}}\otimes\boldsymbol{\bar{3}}=
\boldsymbol{\bar{6}}\oplus\boldsymbol{3}$, which is graphically
reflected by the two tableaux
\begin{center}
\begin{picture}(40,4)(0,0.5)
\linethickness{0.3pt}
\multiput(2,1)(1,0){5}{\line(0,1){3}}
\put(7,2){\line(0,1){2}}
\multiput(2,2)(0,1){3}{\line(1,0){5}}
\put(2,1){\line(1,0){1}}
\put(4,1){\line(1,0){2}}
\multiput(3.5,1.5)(3,0){2}{\circle*{0.4}}
\put(3.1,-0.2){\makebox(1,1){$a_1$}}
\put(4.6,-0.2){\makebox(1,1){$<$}}
\put(6.1,-0.2){\makebox(1,1){$a_2$}}

\put(10,2){\makebox(2,1){and always}}

\multiput(15,1)(1,0){5}{\line(0,1){3}}
\multiput(20,1)(0,2){2}{\line(0,1){1}}
\multiput(15,2)(0,1){3}{\line(1,0){5}}
\put(15,1){\line(1,0){1}}
\put(17,1){\line(1,0){3}}
\multiput(16.5,1.5)(3,1){2}{\circle*{0.4}}

\put(22,3.5){$\textstyle a_1=N-3\mu-\frac{5}{2},\quad a_2=N-3\nu-\frac{1}{2},$}
\put(22,0.8){$\displaystyle 0\le\nu\le\mu\le\frac{N-4}{3}$.}
\end{picture}
\end{center}
We note that $a_2-a_1\ge 2$ even if the colorons occupy adjacent columns, and
that the YHWS $\ket{\m\m}$ belongs to the left tableau. In order to determine
the space generated by $\ket{\m\m}$ we have to investigate the tensor product
$V=V(\boldsymbol{\bar{3}},\xi_1)\otimes V(\boldsymbol{\bar{3}},\xi_2)$, where
$\xi_{1,2}=a_{1,2}-(2N+3)/4$, respectively.  By application of
Sec.~\ref{sec:yangianrepsl3}.iii, $V$ is irreducible. As sl$_3$ representation we
find $V\cong\boldsymbol{\bar{6}}\oplus\boldsymbol{3}$, where the
$\boldsymbol{\bar{6}}$ is represented by the left tableau above and the
$\boldsymbol{3}$ by the right tableau.  The spin currents of $\ket{\m\m}$ are
obtained from the Drinfel'd polynomials of $V$, $P_1(u)=1$ and
$P_2(u)=(u-\xi_1)(u-\xi_2)$, they equal the results derived using explicit
wave functions~\cite{SG05epl}. 

There are fundamentally different two-coloron states, namely the ones
represented by tableaux like
\begin{center}
\begin{picture}(22,4.5)(0,0)
\linethickness{0.3pt}
\multiput(0,1)(1,0){6}{\line(0,1){3}}
\multiput(0,3)(0,1){2}{\line(1,0){5}}
\multiput(0,1)(0,1){2}{\line(1,0){1}}
\multiput(2,1)(0,1){2}{\line(1,0){3}}
\multiput(1.5,1.5)(0,1){2}{\circle*{0.4}}
\put(0.5,0){$a_1, a_2$}

\put(9,3.3){$a_1=N-3\mu+\frac{1}{2},\;a_2=N-3\mu-\frac{1}{2}$,}
\put(9,0.5){$\displaystyle 0\le \mu\le\frac{N-1}{3}$.}
\end{picture}
\end{center}
We have chosen the SMN's to satisfy $a_1=a_2+1$, which is supported by the
following consideration: We rewrite the momentum spacing as $\Delta
p=p_2-p_1=2\pi(g+\ell)/N$, $\ell\in\mathbb{N}_0$, where $g$ denotes the
statistical parameter of the colorons.  With the assignment $a_1=a_2+1$ we
obtain (we keep the relations $p_i\leftrightarrow a_i$) $g=-1/3$ for the
preceding tableau.  Moreover, the momentum spacings for the left two-coloron
tableau above are also given by $p_2-p_1=2\pi(-1/3+\ell)/N$ if $\ell$ takes
the values $\ell\ge 1$. Hence, we obtain $g=-1/3$ for all two-coloron states
where the SU(3) spins of the colorons are coupled antisymmetrically, \ie all
states represented by tableaux where the two colorons occupy different rows.
The finding $g=2/3$ for colour-polarised colorons (in general symmetrically
coupled) and $g=-1/3$ for colorons with different colours (in general
antisymmetrically coupled) is also consistent with what we find by naive state
counting. A negative mutual exclusion statistics was also deduced from the
dynamical spin susceptibility of the SU(3) HSM calculated by
Yamamoto~\ea~\cite{Yamamoto-00prl}, and observed in conformal field theory
spectra analysed by Schoutens~\cite{Schoutens97}.

As a consequence, two colorons occupying the same column transform under the
Y(sl$_3$) representation $V=V(\boldsymbol{\bar{3}},\xi)\otimes
V(\boldsymbol{\bar{3}},\xi-1)$ with $\xi=a_1-(2N+3)/4$. By
Sec.~\ref{sec:yangianrepsl3}.iii, $V$ is reducible and the irreducible
subrepresentation $W$ does not contain the YHWS of $V$ (which is
$\ket{\m\m}$).  As sl$_3$ representation we have $W\cong\boldsymbol{3}$, \ie
the colorons are coupled antisymmetrically.  Hence, if the individual coloron
momenta satisfy $|p_2-p_1|=2\pi/3N$, we deduce with the choice $a_1=a_2+1$ and
the requirement of irreducibility under Y(sl$_3$) transformations that only
the sl$_3$ representation $\boldsymbol{3}$ exists in the spectrum.  This was
also found heuristically in the numerical study of the spectrum of the
HSM~\cite{GS07}, and is consistent with similar results for conformal
field theories~\cite{BouwknegtSchoutens96}.

Furthermore, it is shown in \ref{sec:yangianappsc} that the proper Y(sl$_3$)
subrepresentation $W$ of $V(\boldsymbol{\bar{3}},\xi)\otimes
V(\boldsymbol{\bar{3}},\xi-1)$ is explicitly given by
\begin{equation}
\textstyle W=V\left(\boldsymbol{3},\xi-\frac{1}{2}\right).
\label{eq:yangianreptwocoloronsanti}
\end{equation}
This means that $W$ is a highest weight representation with YHWS
$\ket{\b}\propto\ket{\m\c}-\ket{\c\m}$ (see Fig.~\ref{fig:weightdiagrams}).
We will see below that (\ref{eq:yangianreptwocoloronsanti}) is necessary and
sufficient to built up the complete Hilbert space of the SU(3) HSM with
many-coloron states and the restrictions imposed by the fractional statistics
through the requirement of irreducibility under Y(sl$_3$) transformations.

At this point we wish to underline that fractional statistics in SU($n$) spin
chains cannot be implemented by the requirement of a definite transformation
law under permutations of the spinons (a one-dimensional representation of the
symmetric group), as in specific situations only the antisymmetric spin
representations exist in the spinon Hilbert space, whereas in other situations
only the symmetric representations exist.

\subsection{Three-coloron states}\label{sec:yangianthreecolorons}

If three colorons are present, there are three different cases to be
investigated.  In the first case the SMN's $a_{1,2,3}$ satisfy $a_j-a_i\ge 2$,
$i<j$. The tensor product $V(\boldsymbol{\bar{3}},\xi_1)\otimes
V(\boldsymbol{\bar{3}},\xi_2)\otimes V(\boldsymbol{\bar{3}},\xi_3)$ is
irreducible, hence we find as sl$_3$ representation
$\boldsymbol{\bar{3}}\otimes\boldsymbol{\bar{3}}\otimes\boldsymbol{\bar{3}}=
\boldsymbol{\bar{10}}\oplus\boldsymbol{8}\oplus\boldsymbol{8}
\oplus\boldsymbol{1}$, which is graphically represented by the corresponding
tableaux:
\begin{center}
\begin{picture}(21,3.5)(0,1)
\linethickness{0.3pt}
\multiput(0,2)(1,0){4}{\line(0,1){2}}
\multiput(0,2)(0,1){3}{\line(1,0){3}}
\multiput(0.5,1.5)(1,0){3}{\circle*{0.4}}

\put(4,2){\makebox(1,1){$\oplus$}}

\multiput(6,2)(1,0){2}{\line(0,1){2}}
\put(8,1){\line(0,1){3}}
\multiput(9,1)(0,2){2}{\line(0,1){1}}
\multiput(6,2)(0,1){3}{\line(1,0){3}}
\put(8,1){\line(1,0){1}}
\multiput(6.5,1.5)(1,0){2}{\circle*{0.4}}
\put(8.5,2.5){\circle*{0.4}}

\put(10,2){\makebox(1,1){$\oplus$}}

\multiput(12,2)(3,0){2}{\line(0,1){2}}
\multiput(13,1)(1,0){2}{\line(0,1){3}}
\multiput(12,2)(0,1){3}{\line(1,0){3}}
\put(13,1){\line(1,0){1}}
\multiput(12.5,1.5)(2,0){2}{\circle*{0.4}}
\put(13.5,2.5){\circle*{0.4}}

\put(16,2){\makebox(1,1){$\oplus$}}

\multiput(18,2)(3,-1){2}{\line(0,1){2}}
\multiput(19,1)(1,0){2}{\line(0,1){3}}
\multiput(18,2)(0,1){2}{\line(1,0){3}}
\multiput(19,1)(-1,3){2}{\line(1,0){2}}
\multiput(18.5,1.5)(1,1){3}{\circle*{0.4}}
\end{picture}
\end{center}
We have drawn the tableaux for the smallest possible system with $N=6$, the
situation is unchanged if the colorons do not occupy adjacent columns. The
YHWS $\ket{\m\m\m}$ belongs to the representation $\boldsymbol{\bar{10}}$ (the
left-most tableau).

In the second case two of the colorons occupy the same column, whereas the
third coloron is separated by at least one column. Graphically we have
\begin{center}
\begin{picture}(12,3.5)(0,1)
\linethickness{0.3pt}
\put(3,2){\line(0,1){2}}
\multiput(1,1)(1,0){2}{\line(0,1){3}}
\put(0,3){\line(0,1){1}}
\multiput(0,3)(0,1){2}{\line(1,0){3}}
\put(1,2){\line(1,0){2}}
\put(1,1){\line(1,0){1}}
\multiput(0.5,1.5)(0,1){2}{\circle*{0.4}}
\put(2.5,1.5){\circle*{0.4}}

\put(5,2){\makebox(1,1){or}}

\put(8,2){\line(0,1){2}}
\multiput(9,1)(1,0){2}{\line(0,1){3}}
\put(11,3){\line(0,1){1}}
\multiput(8,3)(0,1){2}{\line(1,0){3}}
\put(8,2){\line(1,0){2}}
\put(9,1){\line(1,0){1}}
\multiput(10.5,1.5)(0,1){2}{\circle*{0.4}}
\put(8.5,1.5){\circle*{0.4}}
\end{picture}
\end{center}
The left tableau stands for the sl$_3$ representation containing the YHWS of
the tensor product $V(\boldsymbol{\bar{3}},\xi_1)\otimes
V(\boldsymbol{\bar{3}},\xi_1-1)\otimes V(\boldsymbol{\bar{3}},\xi_2)$, where
$\xi_2-\xi_1\ge 4$. Using (\ref{eq:yangianreptwocoloronsanti}) the irreducible
subrepresentation of $V(\boldsymbol{\bar{3}},\xi_1)\otimes
V(\boldsymbol{\bar{3}},\xi_1-1)$ is
$V\left(\boldsymbol{3},\xi_1-\frac{1}{2}\right)$, hence the remaining tensor
product $V\left(\boldsymbol{3},\xi_1-\frac{1}{2}\right)\otimes
V(\boldsymbol{\bar{3}},\xi_2)$ is irreducible by
Sec.~\ref{sec:yangianrepsl3}.ii. As sl$_3$ representation we find
$\boldsymbol{8}\oplus\boldsymbol{1}$. The similar result is obtained for the
right tableau.  The sl$_3$ representations $\boldsymbol{8}$ are given by the
tableaux above, the corresponding singlets are represented by
\begin{center}
\begin{picture}(12,3.5)(0,1)
\linethickness{0.3pt}
\put(3,1){\line(0,1){2}}
\multiput(1,1)(1,0){2}{\line(0,1){3}}
\put(0,3){\line(0,1){1}}
\multiput(1,1)(0,1){2}{\line(1,0){2}}
\put(0,3){\line(1,0){3}}
\put(0,4){\line(1,0){2}}
\multiput(0.5,1.5)(0,1){2}{\circle*{0.4}}
\put(2.5,3.5){\circle*{0.4}}

\put(5,2){\makebox(1,1){and}}

\put(8,2){\line(0,1){2}}
\multiput(9,1)(1,0){2}{\line(0,1){3}}
\put(11,1){\line(0,1){1}}
\multiput(8,3)(0,1){2}{\line(1,0){2}}
\put(8,2){\line(1,0){3}}
\put(9,1){\line(1,0){2}}
\multiput(10.5,2.5)(0,1){2}{\circle*{0.4}}
\put(8.5,1.5){\circle*{0.4}}
\end{picture}
\end{center}

In the third case all three colorons are close together, and two of them
occupy the same column. Graphically we have
\begin{center}
\begin{picture}(12,3.5)(0,1)
\linethickness{0.3pt}
\put(1,3){\line(0,1){1}}
\multiput(2,2)(1,0){2}{\line(0,1){2}}
\multiput(1,3)(0,1){2}{\line(1,0){2}}
\put(2,2){\line(1,0){1}}
\multiput(1.5,1.5)(0,1){2}{\circle*{0.4}}
\put(2.5,1.5){\circle*{0.4}}

\put(5,2){\makebox(1,1){or}}

\multiput(8,2)(1,0){2}{\line(0,1){2}}
\put(10,3){\line(0,1){1}}
\multiput(8,3)(0,1){2}{\line(1,0){2}}
\put(8,2){\line(1,0){1}}
\multiput(9.5,1.5)(0,1){2}{\circle*{0.4}}
\put(8.5,1.5){\circle*{0.4}}
\end{picture}
\end{center}
For example, the left tableau is translated into the tensor product
$V(\boldsymbol{\bar{3}},\xi)\otimes V(\boldsymbol{\bar{3}},\xi-1)\otimes
V(\boldsymbol{\bar{3}},\xi+1)$.  As in the second case we first construct the
irreducible subrepresentation of the first two factors, which is given by
$V\left(\boldsymbol{3},\xi-\frac{1}{2}\right)$. The remaining tensor product
$V\left(\boldsymbol{3},\xi-\frac{1}{2}\right)\otimes
V(\boldsymbol{\bar{3}},\xi+1)$ is reducible by
Sec.~\ref{sec:yangianrepsl3}.ii, its irreducible subrepresentation is as
sl$_3$ representation given by $\boldsymbol{8}$.  This is reflected in the
fact that no singlet tableau with the same SMN's exists.  The loss of the
singlet in this case was also observed in conformal field theory
spectra~\cite{BouwknegtSchoutens96}.

\subsection{Many-coloron states}

Let us first consider four colorons forming two antisymmetrically coupled
pairs. Hence we have to investigate the tensor product
$V=V(\boldsymbol{3},\xi_1)\otimes V(\boldsymbol{3},\xi_2)$, where the results
of Sec.~\ref{sec:yangianrepsl3}.i apply. If $\xi_2-\xi_1>1$, then $V$ is
irreducible and $V\cong\boldsymbol{6}\oplus\boldsymbol{\bar{3}}$. If, however,
$\xi_2-\xi_1=1$, then $V$ is reducible and its irreducible subrepresentation
is, as sl$_3$ representation, given by $\boldsymbol{6}$. These two situations
are graphically represented by the tableaux
\begin{center}
\begin{picture}(28,4)
\linethickness{0.3pt}
\multiput(6,1)(1,0){4}{\line(0,1){3}}
\put(10,3){\line(0,1){1}}
\multiput(6,3)(0,1){2}{\line(1,0){4}}
\multiput(6,2)(2,0){2}{\line(1,0){1}}
\multiput(6,1)(2,0){2}{\line(1,0){1}}
\multiput(7.5,1.5)(2,0){2}{\circle*{0.4}}
\multiput(7.5,2.5)(2,0){2}{\circle*{0.4}}
\put(0,2){\makebox(3,1){$\xi_2-\xi_1>1:$}}
\put(12,2){\makebox(2,1){and}}

\multiput(16,1)(1,0){4}{\line(0,1){3}}
\put(20,2){\line(0,1){1}}
\put(16,4){\line(1,0){3}}
\put(16,3){\line(1,0){4}}
\put(19,2){\line(1,0){1}}
\multiput(16,2)(2,0){2}{\line(1,0){1}}
\multiput(16,1)(2,0){2}{\line(1,0){1}}
\multiput(17.5,1.5)(2,0){2}{\circle*{0.4}}
\multiput(17.5,2.5)(2,1){2}{\circle*{0.4}}
\end{picture}
\begin{picture}(28,3)(0,1)
\linethickness{0.3pt}
\multiput(6,1)(1,0){2}{\line(0,1){3}}
\multiput(9,1)(1,0){2}{\line(0,1){3}}
\put(8,3){\line(0,1){1}}
\multiput(6,3)(0,1){2}{\line(1,0){4}}
\multiput(6,2)(3,0){2}{\line(1,0){1}}
\multiput(6,1)(3,0){2}{\line(1,0){1}}
\multiput(7.5,1.5)(1,0){2}{\circle*{0.4}}
\multiput(7.5,2.5)(1,0){2}{\circle*{0.4}}
\put(0,2){\makebox(3,1){$\xi_2-\xi_1=1:$}}
\put(11.8,2){\makebox(3,1){only.}}
\end{picture}
\end{center}

If more than four colorons are present, the corresponding product
representation of Y(sl$_3$) has to contain more than two fundamental
representations. As the representation theory for Y(sl$_3$) is not known in
the same detail as that of Y(sl$_2$), we have to restrict ourselves to some
illuminating examples.  Let us start with the highest-weight tableau
\begin{center}
\begin{picture}(25,3.5)(0,0)
\linethickness{0.3pt}
\multiput(3,2)(0,1){2}{\line(1,0){4}}
\put(5,1){\line(1,0){1}}
\put(4,1){\line(1,0){1}}
\multiput(3,2)(1,0){5}{\line(0,1){1}}
\multiput(4,1)(1,0){3}{\line(0,1){1}}
\multiput(3.5,1.5)(3,0){2}{\circle*{0.4}}
\multiput(3.5,0.5)(1,0){4}{\circle*{0.4}}

\put(10,2.3){SMN's:}
\put(14.5,2.3){$a_1=\frac{3}{2}$, $a_2=\frac{1}{2}$, $a_3=\frac{5}{2}$,}
\put(14.5,0.3){$a_4=\frac{9}{2}$, $a_5=\frac{13}{2}$, $a_6=\frac{11}{2}$,}
\end{picture}
\end{center}
which stands for the tensor product (we have coupled colorons in the same
column already)
\begin{equation}
\textstyle
V\left(\boldsymbol{3},-\frac{11}{4}\right)\otimes 
V\left(\boldsymbol{\bar{3}},-\frac{5}{4}\right)
\otimes V\left(\boldsymbol{\bar{3}},\frac{3}{4}\right)\otimes 
V\left(\boldsymbol{3},\frac{9}{4}\right).
\end{equation}
Using Sec.~\ref{sec:yangianrepsl3}.ii the first as well as the third tensor
product is reducible, its irreducible subrepresentations are
$V(\boldsymbol{8},\zeta_1)$ and $V(\boldsymbol{8},\zeta_2)$, respectively.  We
have not determined the spectral parameters explicitly, but expect them to
satisfy $-11/4<\zeta_1<-5/4$ and $3/4<\zeta_2<9/4$ (in analogy to the
Y(sl$_2$) case~\cite{ChariPressley90}).  Thus we have $\zeta_2-\zeta_1>2$,
which causes the irreducibility of $V=V(\boldsymbol{8},\zeta_1)\otimes
V(\boldsymbol{8},\zeta_2)$~\cite{Molev02}.  As sl$_3$
representation we find
\begin{equation}
  \boldsymbol{8}\otimes\boldsymbol{8}\;=\;\boldsymbol{27}\oplus
  \boldsymbol{10}\oplus\boldsymbol{\bar{10}}
  \oplus\boldsymbol{8}\oplus\boldsymbol{8}\oplus\boldsymbol{1}. 
\label{eq:yangianhelp2}
\end{equation}
The irreducibility of $V$ is confirmed by inspection of the allowed
tableaux with six colorons and the SMN's given above, which are
\begin{center}
\begin{picture}(34,3.5)(-6,0)
\linethickness{0.3pt}
\multiput(-6,2)(0,1){2}{\line(1,0){4}}
\put(-5,1){\line(1,0){1}}
\put(-4,1){\line(1,0){1}}
\multiput(-6,2)(1,0){5}{\line(0,1){1}}
\multiput(-5,1)(1,0){3}{\line(0,1){1}}
\multiput(-5.5,1.5)(3,0){2}{\circle*{0.4}}
\multiput(-5.5,0.5)(1,0){4}{\circle*{0.4}}

\multiput(0,2)(0,1){2}{\line(1,0){4}}
\put(1,1){\line(1,0){2}}
\put(2,0){\line(1,0){1}}
\multiput(0,2)(1,0){5}{\line(0,1){1}}
\multiput(1,1)(1,0){2}{\line(0,1){1}}
\multiput(2,0)(1,0){2}{\line(0,1){1}}
\multiput(0.5,0.5)(0,1){2}{\circle*{0.4}}
\multiput(3.5,0.5)(0,1){2}{\circle*{0.4}}
\multiput(1.5,0.5)(1,1){2}{\circle*{0.4}}

\put(6,2){\line(1,0){4}}
\multiput(6,3)(1,-2){2}{\line(1,0){3}}
\multiput(6,2)(1,0){4}{\line(0,1){1}}
\multiput(7,1)(1,0){4}{\line(0,1){1}}
\multiput(6.5,0.5)(0,1){2}{\circle*{0.4}}
\multiput(9.5,0.5)(0,2){2}{\circle*{0.4}}
\multiput(7.5,0.5)(1,0){2}{\circle*{0.4}}

\put(12,2){\line(1,0){4}}
\multiput(12,3)(1,-2){2}{\line(1,0){2}}
\multiput(15,3)(-1,-3){2}{\line(1,0){1}}
\multiput(12,2)(4,0){2}{\line(0,1){1}}
\multiput(14,0)(1,0){2}{\line(0,1){3}}
\put(13,1){\line(0,1){2}}
\multiput(12.5,0.5)(0,1){2}{\circle*{0.4}}
\multiput(15.5,0.5)(0,1){2}{\circle*{0.4}}
\multiput(13.5,0.5)(1,2){2}{\circle*{0.4}}

\put(18,2){\line(1,0){4}}
\multiput(18,3)(1,-2){2}{\line(1,0){3}}
\put(20,0){\line(1,0){1}}
\multiput(18,2)(4,-1){2}{\line(0,1){1}}
\multiput(20,0)(1,0){2}{\line(0,1){3}}
\put(19,1){\line(0,1){2}}
\multiput(18.5,0.5)(0,1){2}{\circle*{0.4}}
\multiput(21.5,0.5)(0,2){2}{\circle*{0.4}}
\multiput(19.5,0.5)(1,1){2}{\circle*{0.4}}

\multiput(24,2)(1,-1){2}{\line(1,0){3}}
\multiput(24,3)(2,-3){2}{\line(1,0){2}}
\multiput(24,2)(4,-2){2}{\line(0,1){1}}
\multiput(25,1)(2,-1){2}{\line(0,1){2}}
\put(26,0){\line(0,1){3}}
\multiput(24.5,0.5)(0,1){2}{\circle*{0.4}}
\multiput(27.5,1.5)(0,1){2}{\circle*{0.4}}
\multiput(25.5,0.5)(1,2){2}{\circle*{0.4}}
\end{picture}
\end{center}

A similar example is obtained from the tableau
\begin{center}
\begin{picture}(25,3.5)(0,0)
\linethickness{0.3pt}
\multiput(3,2)(0,1){2}{\line(1,0){4}}
\put(6,1){\line(1,0){1}}
\put(4,1){\line(1,0){1}}
\multiput(4,1)(1,0){4}{\line(0,1){2}}
\put(3,2){\line(0,1){1}}
\multiput(3.5,1.5)(2,0){2}{\circle*{0.4}}
\multiput(3.5,0.5)(1,0){4}{\circle*{0.4}}

\put(10,2.3){SMN's:}
\put(14.5,2.3){$a_1=\frac{3}{2}$, $a_2=\frac{1}{2}$, $a_3=\frac{5}{2}$,}
\put(14.5,0.3){$a_4=\frac{9}{2}$, $a_5=\frac{7}{2}$, $a_6=\frac{11}{2}$}
\end{picture}
\end{center}
We have to determine the irreducible subrepresentation of the tensor
product
\begin{equation}
\textstyle
V\left(\boldsymbol{3},-\frac{11}{4}\right)\otimes 
V\left(\boldsymbol{\bar{3}},-\frac{5}{4}\right)
\otimes V\left(\boldsymbol{3},\frac{1}{4}\right)\otimes 
V\left(\boldsymbol{\bar{3}},\frac{7}{4}\right).
\end{equation}
As before, we obtain as intermediate result $V=V(\boldsymbol{8},\zeta_1)\otimes
V(\boldsymbol{8},\zeta_2)$, but this time the spectral parameters satisfy
$-11/4<\zeta_1<-5/4$ and $1/4<\zeta_2<7/4$. In particular, they are separated
by $1/2$ less than in the preceding example.  Inspection of the allowed
tableaux with six colorons and the given SMN's, which are 
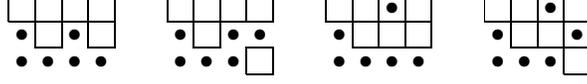
\begin{center}
\begin{picture}(22,3.5)(-6,0)
\linethickness{0.3pt}
\multiput(-6,2)(0,1){2}{\line(1,0){4}}
\put(-5,1){\line(1,0){1}}
\put(-3,1){\line(1,0){1}}
\multiput(-5,1)(1,0){4}{\line(0,1){2}}
\put(-6,2){\line(0,1){1}}
\multiput(-5.5,1.5)(2,0){2}{\circle*{0.4}}
\multiput(-5.5,0.5)(1,0){4}{\circle*{0.4}}

\multiput(0,2)(0,1){2}{\line(1,0){4}}
\put(1,1){\line(1,0){1}}
\multiput(3,0)(0,1){2}{\line(1,0){1}}
\multiput(0,2)(1,0){5}{\line(0,1){1}}
\multiput(1,1)(1,0){2}{\line(0,1){1}}
\multiput(3,0)(1,0){2}{\line(0,1){1}}
\put(0.5,1.5){\circle*{0.4}}
\multiput(0.5,0.5)(1,0){3}{\circle*{0.4}}
\multiput(2.5,1.5)(1,0){2}{\circle*{0.4}}

\put(6,2){\line(1,0){4}}
\put(7,1){\line(1,0){3}}
\put(6,3){\line(1,0){2}}
\put(9,3){\line(1,0){1}}
\multiput(6,2)(1,0){5}{\line(0,1){1}}
\multiput(7,1)(1,0){4}{\line(0,1){1}}
\multiput(6.5,0.5)(0,1){2}{\circle*{0.4}}
\multiput(8.5,0.5)(0,2){2}{\circle*{0.4}}
\multiput(7.5,0.5)(2,0){2}{\circle*{0.4}}

\put(12,2){\line(1,0){4}}
\multiput(12,3)(1,-2){2}{\line(1,0){2}}
\multiput(15,3)(0,-1){4}{\line(1,0){1}}
\multiput(12,2)(4,0){2}{\line(0,1){1}}
\put(16,0){\line(0,1){1}}
\put(15,0){\line(0,1){3}}
\multiput(13,1)(1,0){2}{\line(0,1){2}}
\multiput(12.5,0.5)(0,1){2}{\circle*{0.4}}
\multiput(14.5,0.5)(1,1){2}{\circle*{0.4}}
\multiput(13.5,0.5)(1,2){2}{\circle*{0.4}}
\end{picture}
\end{center}
shows that the irreducible subrepresentation of $V$ is, as representation of
sl$_3$, given by
$\boldsymbol{27}\oplus\boldsymbol{10}\oplus\boldsymbol{\bar{10}}
\oplus\boldsymbol{8}$.  The difference to (\ref{eq:yangianhelp2}) implies that
$V$ is reducible.  Physically, the fractional statistics of the colorons
restricts the allowed SU(3) representations more than above, as the individual
coloron momenta are closer together.

As final example consider the six-coloron states with SMN's $a_1=3/2$,
$a_2=7/2$, $a_3=5/2$, $a_4=9/2$, $a_5=7/2$, and $a_6=11/2$. For this set of
SMN's there exist only the two tableaux
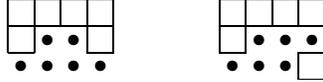
\begin{center}
\begin{picture}(12,3)(0,0.2)
\linethickness{0.3pt}
\multiput(0,2)(0,1){2}{\line(1,0){4}}
\multiput(0,1)(3,0){2}{\line(1,0){1}}
\multiput(0,2)(1,0){5}{\line(0,1){1}}
\multiput(0,1)(1,0){2}{\line(0,1){1}}
\multiput(3,1)(1,0){2}{\line(0,1){1}}
\multiput(1.5,1.5)(1,0){2}{\circle*{0.4}}
\multiput(0.5,0.5)(1,0){4}{\circle*{0.4}}

\multiput(8,2)(0,1){2}{\line(1,0){4}}
\put(8,1){\line(1,0){1}}
\multiput(11,0)(0,1){2}{\line(1,0){1}}
\multiput(8,2)(1,0){5}{\line(0,1){1}}
\multiput(8,1)(1,0){2}{\line(0,1){1}}
\multiput(11,0)(1,0){2}{\line(0,1){1}}
\multiput(9.5,1.5)(1,0){3}{\circle*{0.4}}
\multiput(8.5,0.5)(1,0){3}{\circle*{0.4}}
\end{picture}
\end{center}
\ie the irreducible subrepresentation of 
\begin{equation}
\textstyle
V\left(\boldsymbol{\bar{3}},-\frac{9}{4}\right)\otimes 
V\left(\boldsymbol{3},-\frac{3}{4}\right)
\otimes V\left(\boldsymbol{3},\frac{1}{4}\right)\otimes 
V\left(\boldsymbol{\bar{3}},\frac{7}{4}\right),
\label{eq:yangianmanycoloronsproduct3}
\end{equation}
should be given by $\boldsymbol{27}\oplus\boldsymbol{10}$.

The general scheme for SU(3) works as follows. An $m$-coloron YHWS is
represented by a tableau in which the colorons sit at the bottom of
the columns. First, we couple the colorons in the same column, \ie we
construct the representations $V(\boldsymbol{3},\zeta)$, where $\zeta$
is determined using (\ref{eq:yangianxidefinitionsu3}) and
(\ref{eq:yangianreptwocoloronsanti}). The remaining colorons transform
under $V(\boldsymbol{\bar{3}},\xi)$ with $\xi$ given by
(\ref{eq:yangianxidefinitionsu3}). The space generated under the
action of Y(sl$_3$) by the YHWS is the irreducible subrepresentation
$W$ of the tensor product
\begin{equation}
V=\,\bigotimes_{i=1}^{m'} V\left(\boldsymbol{x}_i,\xi_i\right),\quad 
\xi_1<\xi_2<\ldots<\xi_{m'},
\label{eq:yangiangeneralsu3product}
\end{equation}
where $\boldsymbol{x}_i$ denotes either $\boldsymbol{3}$ or
$\boldsymbol{\bar{3}}$, and $m'$ is the number of occupied columns in the
tableau (the number of isolated colorons plus the number of coloron pairs).
As sl$_3$ representation, $W$ is given by all tableaux with $m$ colorons
possessing the corresponding SMN's. 

To sum up, colorons transform under the Y(sl$_3$) representation
$V(\boldsymbol{\bar{3}},\xi)$, where the spectral parameter $\xi$ is directly
connected to the individual coloron momentum.  The space of $m$ colorons with
momenta $p_1,\dots,p_m$ is generated by the YHWS of
(\ref{eq:yangiangeneralsu3product}) as explained above. The restrictions on
the SU(3) content of this space are due to the fractional statistics of the
colorons.  From a mathematical point of view the tableau formalism provides an
algorithm to derive the sl$_3$ content of the irreducible subrepresentation of
a tensor product of fundamental Y(sl$_3$) representations
(\ref{eq:yangiangeneralsu3product}) with increasing spectral parameters.  As a
by-product, this yields an irreducibility criterion for tensor products of the
form (\ref{eq:yangiangeneralsu3product}).

\section{Conclusion}

In conclusion, we have investigated the relation between the spinon
excitations of the Haldane--Shastry model and its Yangian symmetry.  Each
individual spinon transforms under the fundamental representation of the
Yangian. The associated spectral parameter is directly proportional to its
momentum.  We have obtained a generalised Pauli principle which states that
the spinon Hilbert space is built up by the irreducible subrepresentations of
tensor products of these fundamental representations. This enabled us to
derive several restrictions on the total spin of many-spinon states.  Although
the fractional statistics of spinons can be implemented using the
representation theory of the Yangian only for spinons in the Haldane--Shastry
model, we expect the rules governing the allowed values of the total spin of
many-spinon states to be valid for interacting spinons in general spin chains
as well.

\section*{Acknowledgments}

I would like thank Frank G\"ohmann and Martin Greiter for valuable
discussions. This work was mainly carried out at the Institut f\"ur Theorie
der Kondensierten Materie, Universit\"at Karlsruhe.  This work was supported
by the German Research Foundation (DFG) through GK 284, the Center for
Functional Nanostructures (CFN) Karlsruhe and the Deutsche Akademie der
Naturforscher Leopoldina under grant no BMBF-LPD 9901/8-145.

\appendix

\section{Realization of $\boldsymbol{V(\bar{3},\xi)}$ as evaluation
  representation}\label{sec:yangianappev}

Consider the representation $V(\boldsymbol{\bar{3}},\xi)$ with Drinfel'd
polynomials $P_1(u)=1$ and $P_2(u)=u-\xi$, and denote by $\ket{\m}$ its YHWS.
Then we have
\begin{equation}
H_{1,0}\ket{\m}=H_{1,1}\ket{\m}=0,\quad H_{2,0}\ket{\m}=\ket{\m}\!,
\quad H_{2,1}\ket{\m}=\xi\ket{\m}\!,
\end{equation}
and with (\ref{eq:cwblambda8}) we deduce
\begin{equation}
\Lambda^8\ket{\m}=\frac{1}{\sqrt{3}}\left(\xi+\frac{1}{4}\right)\ket{\m}\!.
\label{eq:appyangian1}
\end{equation}
On the other hand we find with (\ref{eq:yangianevdef3}) that
\begin{eqnarray}
\mathrm{ev}_\zeta(\Lambda^8)&=&\;\zeta J^8+
\frac{1}{2\sqrt{3}}\left(J^3J^3-J^8J^8\right)+
\frac{1}{4\sqrt{3}}\left(I^+I^-+I^-I^+\right)\nonumber\\
& &-\frac{1}{8\sqrt{3}}\left(U^+U^-+U^-U^++V^+V^-+V^-V^+\right)\!,
\end{eqnarray}
and hence for the action of $\Lambda^8$ on $\ket{\m}$ in the evaluation
representation $\phi_\zeta$
\begin{equation}
\Lambda^8\ket{\m}=\frac{1}{\sqrt{3}}\left(\zeta-\frac{5}{12}\right)\ket{\m}\!.
\label{eq:appyangian2}
\end{equation}
Comparison of (\ref{eq:appyangian1}) and (\ref{eq:appyangian2}) yields 
$\zeta=\xi+2/3$.

\section{Irreducible subrepresentation of 
  $\boldsymbol{V(\bar{3},\xi)\otimes V(\bar{3},\xi-1)}$}
\label{sec:yangianappsc}

Consider the tensor product $V=V(\boldsymbol{\bar{3}},\xi_1)\otimes
V(\boldsymbol{\bar{3}},\xi_2)$.  $V$ contains a proper Y(sl$_3$)
subrepresentation $W$ isomorphic to $\boldsymbol{3}$ as sl$_3$ representation
if and only if~\cite{ChariPressley96} $\xi_1-\xi_2=1$.  We wish to determine
the Drinfel'd polynomials of $W$. For that we have to evaluate the actions of
$H_{1,1}$ and $H_{2,1}$ on the YHWS
$\ket{\b}=\ket{\m}\otimes\ket{\c}-\ket{\c}\otimes\ket{\m}$, where
$\ket{\b}\in\boldsymbol{3}$ and $\ket{\m},\ket{\c}\in\boldsymbol{\bar{3}}$.

First, we obtain from (\ref{eq:cwblambda3})
\begin{equation}
\Lambda^3\ket{\b}=\frac{1}{2}H_{1,1}\ket{\b}+\frac{1}{8}\ket{\b}\!.
\label{eq:appyangian3}
\end{equation}
On the other hand we find the action of $\Lambda^3$ on $V$ to be
\begin{eqnarray}
\Delta(\Lambda^3)&=&\Lambda^3\otimes 1+1\otimes\Lambda^3-f^{3ab}J^a\otimes
J^b\nonumber\\
&=&\Lambda^3\otimes 1+1\otimes\Lambda^3+\frac{1}{2}\left(I^+\otimes
  I^--I^-\otimes I^+\right)\nonumber\\
& &-\frac{1}{4}\left(U^+\otimes U^--U^-\otimes U^--
V^+\otimes V^-+V^-\otimes V^+\right)\!.
\end{eqnarray}
On each factor of $V$ the action of $\Lambda^3$ is by
(\ref{eq:yangianevdef3})
\begin{equation}
\hspace{-10mm}
\mathrm{ev}_{\xi_{1,2}}(\Lambda^3)=\xi_{1,2} J^3+
\frac{1}{\sqrt{3}}\,J^3J^8
-\frac{1}{8}\left(U^+U^-+U^-U^+-V^+V^--V^-V^+\right)\!,
\end{equation}
especially we get
\begin{equation}
\Lambda^3\ket{\m}=0,\quad\Lambda^3\ket{\c}=
\left(\frac{\xi_{1,2}}{2}+\frac{1}{8}\right)\ket{\c}\!.
\end{equation}
Hence, we find 
\begin{equation}
\hspace{-10mm}
\Delta(\Lambda^3)\ket{\b}=
\left(\frac{\xi_2}{2}+\frac{3}{8}\right)\ket{\m}\otimes\ket{\c}-
\left(\frac{\xi_1}{2}-\frac{1}{8}\right)\ket{\c}\otimes\ket{\m}=
\left(\frac{\xi_1}{2}-\frac{1}{8}\right)\ket{\b}\!.\label{eq:appyangian4}
\end{equation}
The last equality is valid if and only if $\xi_1-\xi_2=1$, \ie when
the Y(sl$_3$) subrepresentation $W$ is indeed isomorphic to
$\boldsymbol{3}$. With (\ref{eq:appyangian3}) we deduce using
$\xi\equiv\xi_1=\xi_2+1$ that
$H_{1,1}\ket{\b}=\left(\xi-1/2\right)\ket{\b}$. As $W$ can be
constructed as evaluation representation using
(\ref{eq:yangianevdef3}) we obtain $\Lambda^8=\Lambda^3/\sqrt{3}$, and
with (\ref{eq:cwblambda8}) as well as (\ref{eq:appyangian4}) we find
$H_{2,1}\ket{\b}=0$.  Hence, the Drinfel'd polynomials of $W$ are
$P_1(u)=u-\left(\xi-1/2\right)$ and $P_2(u)=1$.


\end{document}